%% file: report.tex
\documentclass[journal,twoside,nofonttune]{IEEEtran}
\usepackage{cite}
\usepackage{amsmath,amssymb,amsfonts}
\usepackage{graphicx}
\usepackage{textcomp}
\usepackage{xcolor}
\usepackage{placeins}
\usepackage{bm}
\usepackage{tikz}
\usepackage{dsfont}
\usepackage[framemethod=TikZ]{mdframed}
\usepackage{microtype}

\usepackage{pgfplots}
\pgfplotsset{compat=1.18}
\usepgfplotslibrary{fillbetween}

\usepackage[font=footnotesize]{caption}

\usepackage{algorithm}
\usepackage{algpseudocode}
\newcommand{\Input}{\State\textbf{Input: }}
\newcommand{\Output}{\State\textbf{Output: }}

\usepackage[utf8x]{inputenc} 
\usepackage[english]{babel}

\usepackage{balance}

\usepackage{diagbox}

\newcommand{\reals}{\mathbb R}      
\newcommand{\integers}{\mathbb Z}   
\newcommand{\naturals}{\mathbb N}   

\DeclareMathOperator*{\argmin}{argmin}

\def\A{{\mathbf A}}

\def\B{{\mathbf B}}
\def\C{{\mathbf C}}
\def\D{{\mathbf D}}

\def\v{{\mathbf v}}
\def\x{{\mathbf x}}

\def\y{{\mathbf y}}

\def\r{{\mathbf r}}
\def\Id{\mathbf{Id}}

\usepackage{multirow}
\newcommand{\cred}{\textcolor{red}}

\newcommand{\cblue}{}

\makeatletter
\def\blfootnote{\xdef\@thefnmark{}\@footnotetext}
\makeatother

\begin{document}
    \title{GraphGrad: Efficient Estimation of Sparse Polynomial Representations for General State-Space Models}
    \author{Benjamin~Cox,~\IEEEmembership{Student Member,~IEEE},
        \'Emilie Chouzenoux,~\IEEEmembership{Senior Member,~IEEE}, and\\
	V\'ictor~Elvira,~\IEEEmembership{Senior Member,~IEEE}

}
    \markboth{IEEE TRANSACTIONS ON SIGNAL PROCESSING, VOL. XX, NO. XX, YYYY}
    {Cox, Chouzenoux, \MakeLowercase{\textit{and}}  Elvira: Efficient Estimation of Sparse Polynomial Representations for General State-Space Models}
    \maketitle
    \blfootnote{B.C. acknowledges support from the \emph{Natural Environment Research Council} of the UK through a SENSE CDT studentship (NE/T00939X/1). The work of V. E. is supported by the ARL/ARO under grant W911NF-22-1-0235. \'E.C. acknowledges support from
the European Research Council Starting Grant MAJORIS ERC-2019-STG850925.}
        
    \begin{abstract}
        State-space models (SSMs) are a powerful statistical tool for modelling time-varying systems via a latent state. 
        In these models, the latent state is never directly observed. 
        Instead, a sequence of observations related to the state is available. 
        The SSM is defined by the state dynamics and the observation model, both of which are described by parametric distributions.
        Estimation of parameters of these distributions is a very challenging, but essential, task for performing inference and prediction.
        Furthermore, it is typical that not all states of the system interact.
        We can therefore encode the interaction of the states via a graph, usually not fully connected.
        However, most parameter estimation methods do not take advantage of this feature. 
        In this work, we propose GraphGrad, a fully automatic approach for obtaining sparse estimates of the state interactions of a non-linear SSM via a polynomial approximation.
        This novel methodology unveils the latent structure of the data-generating process, allowing us to infer both the structure and value of a rich and efficient d of a general SSM. 
        Our method utilises a differentiable particle filter to optimise a Monte Carlo likelihood estimator.
        It also promotes sparsity in the estimated system through the use of suitable proximity updates, known to be more efficient and stable than subgradient methods.
        As shown in our paper, a number of well-known dynamical systems can be accurately represented and recovered by our method, providing basis for application to real-world scenarios.
    \end{abstract}

    \section{Introduction}
        State-space models (SSMs) are a powerful tool for describing systems that evolve in time.
        SSMs are utilised in many fields, such as target tracking, \cite{wang2017survey}, epidemiology \cite{patterson2017statistical}, ecology \cite{newman2023state}, finance \cite{virbickaite2019particle}, and meteorology \cite{clayton2013operational}. 
        SSMs represent a dynamical system via an unobserved hidden state and a series of related observations.
        The objective is then to estimate the hidden state, conditional on the series of observations.
        Estimation of the state using only observations available at the time the state occurs is called the filtering problem, while estimation of the state using the entire observation series, is known as the smoothing problem.
        In order to solve either of these problems, one must know both the conditional densities of the hidden state and of the observation.
        
        When the state transition and observation model are linear and admit Gaussian densities, a linear-Gaussian SSM is obtained, and one can solve the filtering and smoothing problems exactly using the Kalman filter \cite{kalman1960new} and RTS smoother \cite{rauch1965maximum} respectively.
        If the state dynamics or observation model are not linear, then we have a non-linear state-space model (NLSSM), and the Kalman filter cannot be applied without simplifying the model.
        In this case we can use Gaussian approximations, such as the Unscented Kalman filter \cite{wan2000unscented} or the extended Kalman filter, although in many cases a Gaussian approximation is not sufficiently rich to capture the behaviour of the system.
        In such cases, more accurate results might be obtained by a particle filter, which approximates the posterior density of the state via a series of importance weighted Monte Carlo samples \cite{doucet2009tutorial}.
        However, all of these techniques require knowing the form of the transition and observations models, and for all parameters therein to be known.
        It is common for the parameters to be unknown, and therefore they must be estimated.
        Furthermore, in many cases, the form of the model is not known, and therefore we must impose a form before we can estimate anything.
        \cblue{In the case of dynamical systems where we observe the state directly, methods such as SINDy \cite{brunton2016discovering} can be used to recover the system dynamics in the presence of unknown dynamics.
        However these methods cannot be utilised for incompletely observed noisy dynamics, as is the case in state-space models.}
        
        Estimating the parameters of a general non-linear SSM is a difficult task, even when the model is known.
        One challenge is that the exact likelihood is usually intractable for general non-linear models, and we must use a stochastic estimate thereof. 
        A further problem is that we cannot efficiently compute gradients of the likelihood, and hence we cannot easily compute the maximum likelihood estimator, nor utilise standard modern sampling schemes such as Hamiltonian Monte Carlo.
        Thanks to the recent advent of differentiable particle filters \cite{karkus2018particle, corenflos2021differentiable, scibior2021differentiable}, it recently became possible to obtain the gradient of the estimated likelihood of model parameters, and hence became much simpler to fit them using, e.g., maximum likelihood, assuming the form of the model is known.

        In dynamical systems, many phenomena occur with rates proportional to a polynomial function of the state dimensions.
        For example, in population modelling, birth-immigration-death-emigration models are often used, within which the dynamics of the rates are dependent on 2nd degree polynomials of the state-space.
        Furthermore, the well known Lorenz 63 \cite{lorenz1963deterministic} and Lorenz 96 \cite{lorenz1996predictability} models both have rates of change described by 2nd degree polynomials, and are known to be highly chaotic systems of equations.
        We can therefore see that polynomial systems can describe complex phenomena, and hence a crucial problem is to learn the coefficients of a polynomial approximation to the transition function of a NLSSM.

        \vspace{1em}\noindent{\textbf{Contribution}.} In this paper, we propose GraphGrad, a method to model and learn the transition distribution of a non-linear SSM via a polynomial approximation of the state.\footnote{A limited version of this work was presented by the authors in the conference paper \cite{cox2025learning}, which presented a version of this work using automatic differentiation through the $L1$ norm regularisation, and which does not utilise observation batching to stabilise the learning. 
        \cblue{We propose here a more methodologically advanced method, which utilises the proximal operator for applying the $L1$ regularisation, thereby making the method more computationally efficient as it is faster to converge, and have added batch-stabilised learning for long observation series, allowing the method to be applied to large systems without careful initialisation. 
        Furthermore, we include extensive methodological discussion, and provide practical guidance for practitioners. }
        Finally, we include a large number of numerical experiments, which were not present in the prior work.}

            \begin{itemize}
                \item The proposed approximation can represent a rich class of systems, as many dynamical systems are described by a series of differential equations that are polynomial functions of the states. 
                The resulting systems are interpretable, with the interactions between variables having interpretation similar to the rate terms in a dynamical system.
                \item Our method uses a differentiable particle filter, allowing us to use a first order optimisation scheme to perform parameter estimation, improving robustness whilst decreasing computation time. 
                Furthermore, we promote sparse systems by the use of a proximity update, thereby increasing interpretability, and improving stability compared to naive subgradient updates of a penalised loss.
                \item GraphGrad is unsupervised, as we optimise the (penalised) parameter log-likelihood, and therefore do not require knowledge of the underlying hidden state to be trained.
                \item GraphGrad is fast and efficient to evaluate, and provides excellent performance in challenging scenarios, even when the underlying system is not of polynomial form, i.e., under model mismatch.
            \end{itemize}

        \vspace{1em}\noindent{\textbf{Structure}.} In Section~\ref{sec:background}, we present the underlying particle filtering methodology that we build upon, and introduce the differentiable particle filter, as well as the notation used throughout this paper.
        We present the method in Section~\ref{sec:sparse_est}, and discussion in Section~\ref{sec:discussion}.
        We present several numerical experiments in Section~\ref{sec:numerics}, and conclude in Section~\ref{sec:conclusion}.

    \section{Background}
    \label{sec:background}
    
    \subsection{Particle filtering}
    \label{sec:pf}
        The general SSM can be described by
            \begin{align}
                \begin{split}
                    \x_t &\sim p(\x_{t}|\x_{t-1}; \bm\theta),\\
                    \y_t &\sim p(\y_t|\x_t; \bm\theta),
                \end{split} \label{eq:ssm}
            \end{align}
        where $t = 1,\dots, T$ denotes discrete time, $\x_t \in \reals^{N_x}$ is the state of the system at time $t$, $\y_t \in \reals^{N_y}$ is the observation at time $t$, $p(\x_t | \x_{t-1}; \bm\theta)$ is the density of the hidden state $\x_t$, given the previous state $\x_{t-1}$, $p(\y_t | \x_{t}; \bm\theta)$ is the density of the observation $\y_t$ given the hidden state $\x_t$, and $\bm\theta$ is a set of model parameters.
        The initial value of the hidden state is distributed $\x_0 \sim p(\x_0|\bm\theta)$.
        The state sequence $\x_{0:T}$ is typically hidden, whilst the related sequence of observations $\y_{1:T}$ is known.
    
        Filtering methods aim at estimating the hidden state at time $t$, denoted $\x_t$, typically utilising the posterior density function of the state conditional on the observations up to time $t$, denoted $\y_{1:t}$. 
        Particle filters approximate this density, $p(\x_t | \y_{1:t}; \bm\theta)$, using a set of $K$ Monte Carlo samples (particles) and their associated weights, $\{ \x_{1:T}^{(k)}, \widetilde{w}_{1:T}^{(k)} \}_{k=1}^K$. 
        The posterior density can then be approximated by
            \begin{equation}
                p(\x_t | \y_{1:t}; \bm\theta) \approx \sum_{k=1}^K \widetilde{w}_t^{(k)} \delta_{\x_t^{(k)}}.
            \end{equation}
        A commonly used particle filtering method is the sequential importance resampling algorithm, given in Alg.~\ref{alg:sir_pf}. 
        At every time-step $t$, the $K$ particles and normalised weights, $\{ \x_{1:T}^{(k)}, \overline{w}_{1:T}^{(k)} \}_{k=1}^K$, are calculated. 
        First, we perform the resampling step (line \ref{step:resampling}), which generates $K$ samples, sampling $\x_{t-1}^{(k)}, \ k = 1, \dots, K$, with probability $\overline{w}_{t-1}^{(k)}$. 
        The resampling step is vital to avoid the degeneracy of the filter, i.e., to ensure diversity in the particle set and obtain more accurate approximations of the posterior distribution, $p(\x_t | \y_{1:t}; \bm\theta)$. 
        Next, $K$ particles $\x_t^{(k)}$, $k = 1, \dots, K$, are drawn from the proposal distribution $\pi(\x_t | \x_{t-1}, \y_t; \bm\theta)$ (line \ref{step:draw}). 
        Finally, we incorporate the observation $\y_t$, which is done via the particle weights, given by ${w}_t^{(k)}$, $k = 1, \dots, K$, in line \ref{step:weights}, and the normalised weights (line \ref{step:normweights}). 
            \begin{algorithm}[ht]
            \footnotesize
                \caption{Sequential importance resampling (SIR) particle filter}
                \label{alg:sir_pf}
                \begin{algorithmic}[1]
                    \Input Observations $\y_{1:T}$, parameters $\bm\theta$.
                    \Output Hidden state estimates $\x_{1:T}$, particle weights $w_{1:T}$.
                    \State Draw $\x_0^{(k)} \sim p(\x_0|\bm\theta)$, for $k = 1,\dots,K$.
                    \State Set $\widetilde{w}_0^{(k)} = \overline{w}_0^{(k)} = 1/K$, for $k = 1,\dots,K$.
                    \For{$t = 1,\dots,T$ and $k = 1,\dots,K$}
                        \State Sample $a_{t}^{(k)} \sim \mathrm{Categorical}(\overline{w}_{t-1})$. \label{step:resampling}
                        \State Set $\widetilde{w}_{t-1}^{(k)} = 1/K$.
                        \State Sample $\x_t^{(k)} \sim \pi(\x_t|\x_{t-1}^{(a_{t}^{(k)})}, \y_{t}; \bm\theta)$. \label{step:draw}
                        \State Compute $w_t^{(k)} = \frac{p(\y_t|\x_t^{(k)}; \bm\theta)p(\x_t^{(k)}|\x_{t-1}^{(a_{t}^{(k)})}; \bm\theta)}{\pi(\x_t|\x_{t-1}^{(a_{t}^{(k)})}, \y_{t}; \bm\theta)}$. \label{step:weights}
                        \State Compute $\overline{w}_t^{(k)} = \widetilde{w}_{t-1}^{(k)}w_t^{(i)}/\sum_{k=1}^K\widetilde{w}_{t-1}^{(k)}w_t^{(k)}$. \label{step:normweights}
                    \EndFor
                \end{algorithmic}
            \end{algorithm}
    
    \subsection{Parameter estimation in state-space models}
    \label{sec:par_est_ssm}

        In many problems of interest, the parameter $\bm\theta$ is not known, and must be estimated.    
        The posterior distribution of $\bm\theta$ in the SSM can be factorised as $p(\bm\theta|\mathbf{y}_{1:T}) \propto p(\bm\theta)p(\mathbf{y}_{1:T}|\bm\theta)$,
        where $\bm\theta$ is the parameter of interest, $p(\bm\theta)$ is the prior distribution of the parameter $\bm\theta$, and $p(\mathbf{y}_{1:T}|\bm\theta)$ is given by 
            \begin{equation}
            \label{eq:gen_ssm_ll}
                p(\mathbf{y}_{1:T}|\bm\theta) = \prod_{t=1}^{T} p(\mathbf{y}_t|\mathbf{y}_{1:t-1}; \bm\theta),
            \end{equation}
        where $p(\mathbf{y}_1|\mathbf{y}_{1:0}; \bm\theta) := p(\mathbf{y}_1|\bm\theta)$ \cite{sarkka2013bayesian}.
        The prior distribution $p(\bm\theta)$ encodes our pre-existing beliefs as to the value and structure of the parameter $\bm\theta$, and can be used for regularisation, e.g., by the Lasso \cite{tibshirani1996regression}.
        
        There are many methods to estimate parameters $\bm\theta$ given its posterior density function $p(\bm\theta|\mathbf{y}_{1:T})$. 
        We can broadly classify these parameter estimation methods as point estimation methods and distributional methods. 
        Point estimation methods provide a single estimate that is, in some defined way, the optimal value.
        An example of a point estimation method is the maximum-a-posteriori (MAP) estimator, that defines the optimal value of $\bm\theta$ as the one that maximises the posterior density $p(\bm\theta|\mathbf{y}_{1:T})$.
        The method proposed in this work yields a MAP estimator.
        
        In the case of a linear-Gaussian SSM, we have closed form methods for the MAP estimator assuming a diffuse prior \cite{sarkka2013bayesian}, and efficient methods for obtaining the MAP estimator for sparsity promoting priors \cite{elvira2022graphical}.
        Distributional methods estimate the posterior distribution of the parameter, with common methods being importance sampling \cite{tokdar2010importance}, Markov chain Monte Carlo \cite{andrieu2010particle}, and variational inference \cite{blei2017variational}.
        For the linear-Gaussian SSM, we can utilise reversible jump Markov chain Monte Carlo to obtain an estimate of the distribution of sparsity \cite{cox2023sparse}, however no similar method exists for the general SSM.
        Our proposed method in this work is a point estimation method.
        
        For general SSMs, the likelihood of the parameter $\bm\theta$, necessary to compute the posterior and hence to design a procedure to maximise it, cannot be obtained in closed form.
        \cblue{Let $\nu_t^{(k)} = w_t^{(k)} \widetilde{w}_{t-1}^{(k)}$ be the adjusted importance weight.}
        We can estimate \cblue{the parameter likelihood} using the particle filter, by the Monte Carlo estimate
            \begin{align}
                \begin{split}
                    \label{eq:likelihood}
                    p(\bm\theta|\y_{1:T}) &\propto p(\bm\theta)p(\mathbf{y}_{1:T}|\bm\theta),\\ &\approx 
                    p(\bm\theta) \prod_{t=1}^{T} \left(\sum_{k=1}^{K} \nu_t\right),
                \end{split}
            \end{align}
        where $w_t^{(k)}$ and $\widetilde{w}_{t-1}^{(k)}$ are the weights of the particle filter as in Alg.~\ref{alg:sir_pf} \cite[Chapter 12]{sarkka2013bayesian}.
        Note that the weights are dependent on the parameter $\bm\theta$ through their computation in Alg.~\ref{alg:sir_pf}.
        Using these weights, we construct an estimate of the log-likelihood by
            \begin{align}
                \begin{split}
                \label{eq:log_likelihood}
                \log(p(\bm\theta|\y_{1:T})) + c &= \log(p(\bm\theta)) + \log(p(\mathbf{y}_{1:T}|\bm\theta)),\\
                &\approx \log(p(\bm\theta)) + \sum_{t=1}^{T} \log\left(\sum_{k=1}^{K} \nu_t^{(k)}\right),
                \end{split}
            \end{align}
        where the constant $c$ results from the proportionality in Eq.~\eqref{eq:likelihood}, and we note that if resampling occurred at time $t-1$, which is always the case if following Alg.~\ref{alg:sir_pf}, we have $\widetilde{w}_{t-1}^{(k)} = 1/K$, \cblue{and hence $\nu_t^{(k)} = w_t^{(k)}$ due to the weights being normalised}.
        Note that, in practice, log-weights are used for numerical stability, and we compute $\sum_{k=1}^{K} w_t^{(k)} \widetilde{w}_{t-1}^{(k)}$ using a logsumexp re\cred{parameterisation}.
    
    \subsection{Differentiable particle filter}
    \label{sec:diff_pf}
            \begin{algorithm}[ht]
            \footnotesize
                \caption{Stop-gradient differentiable particle filter (DPF) \cite{scibior2021differentiable}}
                \label{alg:sgd_dpf}
                \begin{algorithmic}[1]
                    \Input Observations $\y_{1:T}$, parameters $\bm\theta$.
                    \Output Hidden state estimates $\x_{1:T}$, particle weights $w_{1:T}$.
                    \State Draw $\x_0^{(k)} \sim p(\x_0|\bm\theta)$, for $k = 1,\dots,K$.
                    \State Set $\widetilde{w}_0^{(k)} = \overline{w}_0^{(k)} = 1/K$, for $k = 1,\dots,K$.
                    \For{$t = 1,\dots,T$ and $k = 1,\dots,K$}
                        \State Sample $a_{t}^{(k)} \sim \mathrm{Categorical}(\bot(\overline{w}_{t-1}))$. \label{step_dpf:resample}
                        \State Set $\widetilde{w}_t^{(k)} = \frac{1}{K} \ \overline{w}_{t-1}^{a_{t}^{(k)}} / \bot(\overline{w}_{t-1}^{a_{t}^{(k)}})$. \label{step_dpf:grad_push}
                        \State Sample $\x_t^{(k)} \sim \pi(\x_t|\x_{t-1}^{a_{t}^{(k)}}, \y_{t}; \bm\theta)$. \label{step_dpf:draw}
                        \State Compute $w_t^{(k)} = \frac{p(\y_t|\x_t^{(k)}; \bm\theta)p(\x_t^{(k)}|\x_{t-1}^{a_{t}^{(k)}}; \bm\theta)}{\pi(\x_t|\x_{t-1}^{a_{t}^{(k)}}, \y_{t}; \bm\theta)}$. \label{step_dpf:weights}
                        \State Compute $\overline{w}_t^{(k)} = \widetilde{w}_{t-1}^{(k)}w_t^{(i)}/\sum_{k=1}^{K}\widetilde{w}_{t-1}^{(k)}w_t^{(k)}$. \label{step_dpf:normweights}  
                    \EndFor
                \end{algorithmic}
            \end{algorithm}

        The outputs of the particle filter, as given in Alg.~\ref{alg:sir_pf}, namely $\{\x^{(k)}_{1:T}, w^{(k)}_{1:T}\}_{k=1}^K$, are not differentiable {with respect to $\bm\theta$} \cite{scibior2021differentiable}, because of the resampling step on line \ref{step:resampling}.
        ~The resampling step requires sampling a multinomial distribution.
        Sampling a multinomial distribution is not differentiable, as an infinitesimal change in the input probabilities can lead to a discrete change in the output sample value \cite{scibior2021differentiable}.
        
        Differentiable particle filters (DPFs) are recently introduced tools that modify the resampling step of the SIR particle filter to be differentiable, and therefore lead to a particle filtering algorithm that is differentiable with respect to $\bm\theta$ \cite{corenflos2021differentiable,chen2023overview,scibior2021differentiable,li2023differentiable}.
        There exist a number of DPF methods, based on various techniques for making the resampling step differentiable. 
        These range from weight retention in soft resampling \cite{karkus2018particle}, to optimal transport \cite{corenflos2021differentiable}. 
        An overview of the DPF and its interplay with deep learning methods can be found in \cite{chen2023overview}, which motivates our choice of differentiable particle filter, and our usage of stochastic gradient descent methods.
        
        In this work, we built upon the DPF approach from \cite{scibior2021differentiable}, which utilises a stop gradient operator to make the resampling step differentiable.
        We summarise this method in Alg.~\ref{alg:sgd_dpf}.
        This algorithm yields gradient estimates with minimal computational overhead, and does not modify the behaviour of the forward pass of the particle filter.

        In Alg.~\ref{alg:sgd_dpf}, at each time-step $t$, we first sample the previous particle set, sampling $a_{t}^{(k)}$, $k = 1, \dots, K$, with probability $\overline{w}_{t-1}$ (line \ref{step_dpf:resample}).
        The value of $a_{t}^{(k)}$ determines the ancestry of the $k$-th particle at time $t$.
        Note that we apply a stop gradient operator to the weights in the sampling method, and therefore do not attempt to propagate gradients through sampling a discrete distribution.
        We then set the weights of all $K$ particles to $1/K$ whilst preserving gradient information in the weights by dividing the weights by themselves, applying a stop-gradient operation to the divisor, and then multiplying by the constant $1/K$ (line \ref{step_dpf:grad_push}).
        The rest of the DPF proceeds as Alg.~\ref{alg:sir_pf}, described in Sec.~\ref{sec:pf}, with the particle sampling, weighting, and normalisation steps unmodified.

        The DPF is particularly useful when estimating parameters, as we can compute the gradients of functions of the likelihood and of the particle trajectories, and therefore compute parameters optimising a chosen loss function.
        In particular, we typically use likelihood-based losses when performing inference where the hidden state is unknown in the training data, and trajectory-based losses when the sequence of hidden states is known for the training data.
        Examples of likelihood-based losses are the negative log-likelihood and the ELBO.
        The mean-square error of the inferred state is a typical trajectory-based loss.
        Note that the trajectory of the hidden states depends on the weights, and therefore the weights must be differentiable even if the loss function is based only on the trajectory.

        Given a loss function based on the weights and/or particles of a differentiable particle filter, we can compute the minimiser using a gradient-based optimisation scheme.
        This is efficient, especially compared to the gradient-free methods previously required, and is more robust to the stochasticity of the likelihood and state estimates, as many gradient schemes are designed with noisy gradients in mind \cite{kingma2014adam, li2019jasper}.

    \subsection{Notation}
        \label{sec:notation}

        We here present our notation, used throughout the paper.
        We denote by $\A_{\cdot, i}$ the $i$-th column vector of matrix $\A$, and by $\A_{i, \cdot}$ the $i$-th row vector of matrix $\A$.
        We denote by $\Id_{n}$ the $n \times n$ identity matrix.

        We denote by $\mathds{1}_{\mathrm{cond}}(x)$ the binary indicator function, which returns $1$ when $x$ satisfies $\mathrm{cond}$, and $0$ otherwise.
        We denote by $\mathrm{sgn}(x)=\mathds{1}_{\geq0}(x)-\mathds{1}_{\leq0}(x)$ the sign function.
        We denote by $\mathrm{abs}(x)$ the absolute value function.
        All three functions can be applied element-wise to a matrix or vector. 

        We denote by $\mathbb{N}_0$ the natural numbers including $0$, and by $\mathbb{N}$ the natural numbers excluding $0$.
        
        We denote by $\mathrm{count}(a, b)$ the number of times the item $a$ occurs in the list $b$, and by $\mathrm{cwr}(S, r)$ the set of all length $r$ combinations of the elements of the set $S$ with replacement, up to reordering.
        For example, we have $\mathrm{count}(1, [1, 2, 2, 3, 1]) = 2$ and $\mathrm{cwr}(\{a, b, c\}, 2) = \{[a,a], [a,b], [a,c], [b,b], [b,c], [c,c]\}$.
        
        We denote by $\bot(x)$ the stop-gradient operator applied to $x$, with properties as defined in \cite{scibior2021differentiable}.

    \section{Sparse estimation of non-linear SSMs}
    \label{sec:sparse_est}

        This section presents our main contribution,
        \cblue{a novel approach for modelling and learning the state transition distribution of a general SSM utilising a polynomial approximant.}
        \cblue{
        Several dynamical systems can be exactly represented as polynomials, such as the chaotic Lorenz 63 and 96 systems \cite{lorenz1963deterministic, lorenz1996predictability} or the Rabinovich-Fabrikant system \cite{rabinovich1979stochastic}, the Lotka-Volterra model and many of its extensions \cite{wangersky1978lotka}, many compartmental epidemiological models \cite{brauer2008compartmental}, and several ODEs resulting from PDE discretisations (such as from the Brusselator model \cite{prigogine1968symmetry} and Oregonator model \cite{field1974oscillations}).
        These systems with different properties and dynamics can all be represented exactly by a polynomial model, demonstrating the wide array of systems that a polynomial approximation can exactly capture.
        }
        
        Our approach builds a polynomial approximation to the transition distribution, parametrised by $\C\in\reals^{N_x \times M}$, a matrix of real numbers encoding polynomial coefficients, to be learnt, and $\D \in \naturals_0^{N_x \times M}$ a fixed (i.e., known) integer matrix of monomial degrees associated with $\C$. 
        
        The positive integer $d$ denotes the fixed maximum degree of our polynomial approximation.
        The number of monomials of degree $d$ in $N_x$ variables is  $M = \sum_{n=0}^{d} \binom{n+N_x-1}{N_x-1}$.
        No other model parameters are assumed unknown, hence $\bm\theta$ is equal in our case to $\C$, and, in particular, $p(\x_{t}|\x_{t-1}; \bm\theta) = p(\x_{t}|\x_{t-1}; \C)$. 
        For example, in the additive zero-mean Gaussian noise case, we have
            \begin{equation}
                \label{eq:poly_eval_early}
                \x_{t} \sim p(\x_{t}|\x_{t-1}, \C) := \mathcal{N}(f(\x_{t-1}, \C; \D), \bm\Sigma_v),
            \end{equation}
        where $\bm\Sigma_v$ is the covariance of the state noise distribution, and, for every $\x = [x_1, x_2, \dots, x_{N_x}]^{\top} \in \mathbb{R}^{N_x}$,
            \begin{equation}
                f(\x, \C; \D) = \sum_{j=1}^{M} \left(\C_{\cdot, j} \cdot \prod_{i=1}^{N_x} x_i ^ {D_{i,j}}\right).
                \label{eq:polynomial}
            \end{equation}
        is our considered polynomial approximation.
        The degree matrix $\D$ is constructed from the number of hidden states $N_x$ and the maximum degree $d$, and is a static parameter.

        The remainder of this section is structured as follows.
        Our polynomial model is thoroughly described in Section \ref{sec:gen_mono}.       
        We then design an approach to learn the coefficient matrix $\C$ using a MAP estimator under a sparsity inducing penalty.
        We next discuss in Section \ref{sec:graphical_matrices} the graphical interpretation of $\C$ and the key role of sparsity. 
        We then use the Lorenz 63 dynamical system as a pedagogical example to better illustrate the usefulness of our model, in Section ~\ref{sec:l63_example_section}.
        We move to the presentation of our approach to estimate $\C$, starting from the problem definition in Section \ref{sec:parameter_estimation}, a single batch algorithm S-GraphGrad in Section \ref{sec:init_alg}, and a practical mini-batch implementation, leading to our final algorithm B-GraphGrad, in Section \ref{sec:graphgrad}.        

    \subsection{Constructing a polynomial approximant}
    \label{sec:gen_mono}

        In order to learn a polynomial approximation to the state transition distribution, let us first define the learnt and static parameters of our polynomials.
        A polynomial is the result of summing a number of monomials, and hence can be constructed as a sum of estimated monomial terms.
        The degree of a monomial is given by the sum of the powers of its constituent terms, with a polynomial having degree equal to the maximum degree of its constituent monomials.
        In our model, we assume a fixed maximum degree $d\in\naturals$ for our polynomial approximation. 

        Once $d$ is set, we construct all length $N_x$ sequences of positive integers that sum to $n \leq d, \, n \in \naturals_0$, resulting in 
            \begin{equation}
                \label{eq:num_monom}
                M = \sum_{n=0}^{d} \binom{n+N_x-1}{N_x-1}
            \end{equation}
        unique sequences. 
        This simple procedure allows us to generate the powers of all monomial terms in a polynomial of degree $d$, that we store in an $N_x \times M$ matrix, denoted $\D$, with the term $D_{i,j}$ corresponding to the power of state dimension $i$ in the $j$-th monomial term. 
        The polynomial expression is then defined by matrix $\C$, following Eq.~\eqref{eq:polynomial}.
        The ordering of the monomials is arbitrary but must be consistent, as it implies the order of the columns of $\C$ and $\D$.
        
        \cblue{In our construction, $d$ must be a non-zero natural number, as we construct polynomials from positive integer powers of the state components. 
        However, the method could easily be extended to utilise rational powers of the state, of the form $1/p, \ p \in \naturals$.
        For example, one could construct an approximant utilising square root terms with maximal polynomial degree $d$ simply by constructing polynomials up to degree $2d$, and then substituting in the square root term.
        More generally, approximants of maximal degree $d$ using terms of power $1/p$ could be utilised by constructing polynomials up to degree $pd$, and then dividing $\D$ by $p$.
        We choose to focus here on the integer power polynomials for notational simplicity, ease of implementation, and the capability to be interpreted in a similar fashion to Taylor approximants.}

    \subsubsection{Generating the degree matrix}
    \label{sec:gen_deg}
        We will now specify the construction of $\D$, the degree matrix. 
        As before, $\D$ is such that $D_{i,j}$ is the power of state dimension $i$ in the $j$-th monomial term.
        For example, if $N_x = 3$, and, for some $j \in \{1,\ldots,M\}$, $\D_{\cdot,j} = [0, 1, 2]$, then the value of the $j$-th monomial term when evaluating the transition of the $i$-th state dimension is $C_{i,j} x_1^0x_2^1x_3^2$, where $C_{i,j}$ is a learnt coefficient.
        The degree matrix $\D$ is static, and hence the same for every state, meaning that all states fit a polynomial of the same maximum degree.

        Our method for construction the degree matrix $\D$ is given in Alg.~\ref{alg:generate_D}, where `$\mathrm{count}$' and `$\mathrm{cwr}$' are defined in Sec.~\ref{sec:notation}.
        Alg.~\ref{alg:generate_D} takes the union of all possible combinations with replacement of the set $\{1, 2, \dots, N_x\}$ of length less than or equal to $d$, denoting by $\mathcal{Q}$ the resulting set.
        We then construct $\D$ by setting each entry $D_{i,j}$ equal to the number of times $i$ occurs in the $j$-th element of $\mathcal{Q}$, for $i \in \{1, \dots, N_x\}$ and $j \in \{1, \dots, M\}$.

            \begin{algorithm}[ht]
            {
                \footnotesize
                \caption{Generating the degree matrix $\D$}
                 \label{alg:generate_D}
                 \begin{algorithmic}[1]
                    \Input State size $N_x$, maximal degree $d$.
                    \Output Matrix $\D\in\reals^{N_x \times M}$ of monomial degrees.
                    \State Compute $\mathcal{Q} = \bigcup\limits_{\delta=0}^{d}\mathrm{cwr}([1, 2, \dots, N_x], \delta)$
                    \State $D_{i,j} = \mathrm{count}(i, \mathcal{Q}_j)$ $\forall \ i \in \{1, \dots, N_x\}, \ j \in \{1, \dots, M\}$. 
                 \end{algorithmic}
            }
            \end{algorithm}

        We observe that $|\mathcal{Q}| = M$.
        Note that the set $\mathcal{Q}$ has no inherent ordering, but we access it by index. 
        We must therefore impose an ordering on the set $\mathcal{Q}$.
        One such ordering is lexicographical ordering.
        To apply this ordering, we first count how many times each number appears in an element of $\mathcal{Q}$. 
        We then order these elements by the number of times $1$ appears, and in case of equality comparing the number of times $2$ appears, and so on until $N_x$.
        Note that the ordering of the degree matrix does not change the properties of the algorithm.
        In this work, for interpretability, we sort monomials in ascending order by degree, then sorting by lexicographical order as a tiebreaker within degree, as this aligns closely with how the degree matrix is generated in Alg.~\ref{alg:generate_D}, by iterating over degrees.

    \subsubsection{Initialising the coefficient matrix and evaluating the polynomial}
    \label{sec:gen_coef}
        Now that $\D $ is constructed, to evaluate the resulting polynomial for each state dimension, we require $\C$, encoding the coefficients of each monomial term in every state dimension.
        These coefficients are our object of inference, and therefore should be stored in a manner admitting efficient evaluation of the polynomial.
        
        We have $M$ monomials for each of the $N_x$ state dimensions, and therefore we propose to store the monomial coefficients in an $N_x \times M$ matrix $\C$, where $C_{i,j}$ corresponds to the coefficient of the $j$-th monomial when computing state $i$.
        This gives us a total of $N_x \cdot M = N_x \sum_{n=0}^{d} \binom{n+N_x-1}{N_x-1}$ parameters to estimate.

        Following usual broadcasting rules, given $\x$, $\D$, and $\C$, we can now evaluate the value of our polynomial at any $\x \in \reals^{N_x}$ by Eq.~\eqref{eq:polynomial}.
        Note that $\prod_{i=1}^{N_x} x_i ^ {D_{i,j}}$ is scalar, with $\C_{\cdot, j} \prod_{i=1}^{N_x} x_i ^ {D_{i,j}}$ being the vector $\C_{\cdot, j}$ multiplied by the scalar $\prod_{i=1}^{N_x} x_i ^ {D_{i,j}}$.
        In effect, $\prod_{i=1}^{N_x} x_i ^ {D_{i,j}}$ evaluates the $j$-th monomial term with coefficient $1$, and this calculation is reused for every state dimension, with $\C_{\cdot, j} \prod_{i=1}^{N_x} x_i ^ {D_{i,j}}$ applying the coefficients, which are unique to each state dimension.
        Once the model is initialised, our goal is to learn the coefficient matrix $\C$, since this matrix, in conjunction with the known fixed degree matrix $\D$, defines the transition density $p(\x_{t}|\x_{t-1}; \bm\theta) = p(\x_{t}|\x_{t-1}; \C)$ in Eq.~\eqref{eq:ssm}, where $\bm\theta$ is the set of learnt parameters, in our case $\C$.

    \subsection{A graphical interpretation of matrices $\C$ and $\D$}
    \label{sec:graphical_matrices}
        Within time series modelling, we can often interpret the driving parameters of a sparse model as a graph encoding the connectivity of the system \cite{elvira2022graphical, cox2023sparse, chouzenoux2023graphit, chouzenoux2023sparse, chouzenoux2024graphical, tan2024backpropagation, chouzenoux2024sparse}.
        This is the case here as well, where we observe that, for $(a,b) \in \{1, \dots, N_x\}^2$, state dimension $b$ affects state dimension $a$ in our estimated dynamics if $(\mathrm{abs}(\C)\D^{\top})_{a,b} \neq 0$, where we note that $\mathrm{abs}(\C)\D^{\top}$ is an $N_x \times N_x$ matrix.
        
        We can interpret this in terms of Granger causality, where we see that including information from state $b$ improves the knowledge on state $a$, and therefore we say that state $b$ Granger-causes state $a$ if $(\mathrm{abs}(\C)\D^{\top})_{a,b} \neq 0$.

        We are therefore able to construct a directed graph \cblue{encoding the network topology} of our estimated system from the matrices $\C$ and $\D$.
        This graph has adjacency matrix $\A$, defined by $\A = \mathds{1}_{\neq0}(\mathrm{abs}(\C)\D^{\top}) \in \{0, 1\}^{N_x \times N_x}$, where we have an edge from node $b$ to node $a$ if $A_{a,b} = 1$, and no edge if $A_{a,b} = 0$.
        If this graph is not fully connected, or equivalently if there exist $(a, b) \in \{1, \dots, N_x\}^2$ such that $A_{a,b} = A_{b,a} = 0$, then some state dimensions do not directly interact in our estimated system.
        \cblue{Note that this graph represents the connectivity of the state space within the system dynamics, and therefore it is distinct from graphical SSMs that perform state estimation over graphs, such as \cite{zambon2023graph}.
        However, the interpretation of the above graph as encoding relationships between state dimensions is the same as that in methods that estimate the state as a graph, or methods that infer network structure as a graph \cite{mateos2019connecting}.}

        We can also interpret our system as a collection of $M$ graphs, $\{\mathcal{G}_j\}_{j=1}^{M}$, where the $j$-th graph $\mathcal{G}_j$ has adjacency matrix $\mathrm{abs}(\C_{\cdot,j})\D_{\cdot,j}^{\top}$. 
        Each $\mathcal{G}_j$ can be interpreted as encoding the connectivity resulting from the $j$-th monomial.
        A sparsity promoting prior on $\C$ also promotes sparsity in these graphs.
        We can therefore interpret our method of estimating $\C$ as estimating multiple interacting sparse graphical models, one for each of the underlying monomials. 
        \cblue{We can then recover the overall connectivity graph by taking a union of the graphs encoding the connectivity of the individual monomials.
        These graphs can be constructed for more general forms of model, such as those discussed in Sec.~\ref{sec:discussion:library}, thereby generalising the sparse graphical interpretation to parametric non-linear model discovery methods.}

        \cblue{
        Once estimated, these graphs can be utilised in a variety of ways.
        For example, it is possible that the system could be broken down into sparsely interacting sub-systems, which is a common structure for real-world systems \cite{watts1998collective}, or, if possible, used to separate the system dynamics into non-interacting systems that can be filtered in parallel, assuming the observation model also allows them to be separated.
        The graph estimate can also be used to determine the most influential state dimensions, as the nodes relating to these will have a higher out degree than nodes relating to dimensions that affect fewer other dimensions.
        These examples are not exhaustive, and many potential uses of this graph interpretation are system specific, for example, estimating the network structure of a power grid if observations are related to such a system.
        }

    \subsection{An example: Lorenz 63}
    \label{sec:l63_example_section}

        We present here a worked example utilising the Lorenz 63 model.  
        The Lorenz 63 model \cite{lorenz1963deterministic} is a popular chaotic oscillator model, with a discrete time variant given by,
            {
            \begin{align}
                \setlength{\arraycolsep}{2.5pt}
                \thinmuskip=1mu
                \medmuskip=2mu plus 2mu minus 2mu
                \thickmuskip=3mu plus 5mu minus 2mu
                \begin{split}
                \label{eq:l63_disc_section}
                    x_{1, t} &= x_{1, t-1} + \Delta t (\sigma(x_{2, t-1} - x_{1, t-1})) + v_{1, t},\\
                    x_{2, t} &= x_{2, t-1} + \Delta t (x_{1, t-1} (\rho - x_{3, t-1}) - x_{2, t-1}) + v_{2, t},\\
                    x_{3, t} &= x_{3, t-1} + \Delta t (x_{1, t-1} x_{2, t-1} - \beta x_{3,t-1}) + v_{3, t}, \\
                    \y_{t} &= \x_{t} + \r_t,
                \end{split}
            \end{align}
            }
        where $\Delta t$ is the time elapsed between observations, $\v_{t} \sim \mathcal{N}(\bm0, \Sigma_v)$ the state noise term, and $\r_{t} \sim \mathcal{N}(\bm0, \Sigma_r)$ the observation noise term, and $\beta, \,\rho, \,\sigma$ are real scalar parameters, often taken as $\beta = 8/3, \ \sigma = 10, \ \rho = 28$. The initial condition $\x_0$ is arbitrary, so long as it is non-zero, and is often taken to be $[1, 0, 0]$.

\begin{figure}[ht]
                \centering
                \begin{minipage}{.22\textwidth}
                    \[
                    \A = \begin{pmatrix}
                        1 & 1 & 0\\
                        1 & 1 & 1\\
                        1 & 1 & 1\\
                    \end{pmatrix},
                    \]
                \end{minipage}
                \begin{minipage}{.22\textwidth}
                    \includegraphics[width=0.85\textwidth]{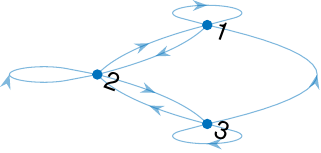}
                \end{minipage}
                \caption{Graph and adjacency matrix encoding the connectivity of the Lorenz 63 system.}
                \label{fig:l63_graph}
            \end{figure}

        We can see that the transition state system in Eq.~\eqref{eq:l63_disc_section} is described by a degree $d=2$ polynomial in $N_x=3$ variables.
        Therefore, using our notations, we have
            \[Q = \{[1,1], [1,2], [1,3], [1], [2,2], [2,3], [2], [3,3], [3], []\}\]
        under lexicographical ordering, and
            \[Q = \{[], [1], [2], [3], [1,1], [1,2], [1,3], [2,2], [2,3], [3,3]\}\]
        under lexicographical-in-degree ordering.
        From the lexicographical-in-degree ordering, the resulting degree matrix $\D$ is
        \begin{equation*}
            \setlength{\arraycolsep}{2.5pt}
            \medmuskip = 1mu 
            \D = \begin{pmatrix}
                0 & 1 & 0 & 0 & 2 & 1 & 1 & 0 & 0 & 0 \\
                0 & 0 & 1 & 0 & 0 & 1 & 0 & 2 & 1 & 0 \\
                0 & 0 & 0 & 1 & 0 & 0 & 1 & 0 & 1 & 2 \\
            \end{pmatrix}.
        \end{equation*}
        Given the above degree matrix $\D$, we can extract from Eq.~\eqref{eq:l63_disc_section} the true coefficient matrix $\C$, 
        \begin{equation*}
            \setlength{\arraycolsep}{2.5pt}
            \medmuskip = 1mu 
            \C = \begin{pmatrix}
                0 & 1-\sigma\Delta t & \sigma\Delta t & 0 & 0 & 0 & 0 & 0 & 0 & 0 \\
                0 & \rho\Delta t & 1 -\Delta t & 0 & 0 & 0 & -\Delta t & 0 & 0 & 0 \\
                0 & 0 & 0 & 1-\beta\Delta t & 0 & \Delta t & 0 & 0 & 0 & 0 \\
            \end{pmatrix},
        \end{equation*}
        which, notably, is sparse with $23$ elements out of $30$ equal to zero.
        We can verify that inputting the above $\C$ and $\D$ into Eq.~\eqref{eq:poly_eval_early} and Eq.~\eqref{eq:polynomial} yields the system given in Eq.~\eqref{eq:l63_disc_section}.
        Furthermore, we construct the adjacency matrix $\A = \mathds{1}_{\neq0}(\mathrm{abs}(\C)\D^{\top})$ from the above $\C$ and $\D$ matrices, yielding the adjacency matrix and associated directed graph given in Fig.~\ref{fig:l63_graph}.

        In this example, the coefficient matrix $\C$ is sparse, while the adjacency matrix $\A$ gives a highly connected graph. 
        Imposing a sparse $\C$ at the estimation stage therefore gives more information than the adjacency matrix $\A$, as we also obtain estimates of the type of interactions that occur between the hidden states.

    \subsection{Parameter estimation}
        \label{sec:parameter_estimation}
        
        We now move to the learning procedure, for the unknown parameter $\bm\theta := \C \in \reals^{N_x \times M}$.
        This is done via a MAP approach, that is by minimising the penalised negative log-likelihood $\ell_R$, given by
            \begin{equation}
                \ell_R(\C| \y_{1:T}, \lambda) = \ell(\C; \y_{1:T}) + \lambda R(\C),
            \end{equation}
        where 
            \begin{equation}
                \ell(\C| \y_{1:T}) = -\log(p(\y_{1:T}|\C)),
            \end{equation}
        and $\lambda R(\C)$ is a sparsity promoting penalty term acting on $\C$, with penalty weight $\lambda>0$.
        Hence, we aim to compute
            \begin{equation}
                \label{eq:objective_function}
                \widehat{\C} = \argmin_{\C \in \reals^{N_x \times M}} \ell_R(\C|\y_{1:T}, \lambda).
            \end{equation}
        We propose to adopt the $L1$ penalty to promote sparsity, given by 
            \begin{equation}
                \label{eq:penalty_R_L1}
                (\forall \C \in \reals^{N_x \times M}) \quad R(\C) = ||\C||_{1}  := \sum_{j=1}^{M}\sum_{i=1}^{N_x}|C_{i,j}|.
            \end{equation}

       We propose to solve Eq.~\eqref{eq:objective_function}, using a first-order optimisation scheme, combining a gradient step on the log-likelihood, and a proximity step over $\lambda R$.
       We will now describe the gradient step, and then describe the proximity step.

    \subsubsection{Estimating the parameter likelihood and its gradient}

        In order to resolve Eq.~\eqref{eq:objective_function} using a first order optimisation scheme, we require to evaluate the negative log-likelihood and its first derivative with respect to $\C$, given the observation series and the SSM.
        We propose to estimate the likelihood through Eq.~\eqref{eq:likelihood}, where we have $\bm\theta := \C$, our parameter of interest.
        We then transform this quantity to the negative log-likelihood through negating the logarithm of the resulting estimate.
        For stability reasons, in practice, log-weights are used, and the log-likelihood is thus computed directly.
        The obtained estimate of the log-likelihood depends on the particle weights, which, in the standard SIR particle filter of Alg.~\ref{alg:sir_pf}, are subject to a non-differentiable resampling step.
        
        Thankfully, we can utilise the DPF approach discussed in Sec.~\ref{sec:diff_pf}, to obtain an estimate of the negative log-likelihood with respect to our parameter $\C$, and the gradient of the negative log-likelihood with respect to $\C$.
        The DPF yields a stochastic estimate of the gradient, which we can use to perform gradient based minimisation of the negative log-likelihood.
        In order to be robust to gradient noise, we propose to rely on first-order updates from the deep learning literature, where stochastic gradients are common due to stochastic input batches.
        In our experiments, we utilise the Novograd optimiser \cite{li2019jasper, ginsburg2019training} to compute our gradient updates for the negative log-likelihood, as it is robust to gradient outliers \cite{ginsburg2019training}.

        We denote the result of the gradient update utilising the negative log-likelihood at iteration $s$, 
            \begin{equation}
                \widetilde{\C}_s = \mathrm{update}_\eta(\C_{s-1}, \nabla\ell(\C_{s-1}|\y_{1:T})),
            \end{equation}
        where $\mathrm{update}_\eta(\A, \nabla\ell(\B|\y_{1:T}))$ is a step of a given minimisation scheme with learning rate $\eta$ applied to $\A$ with gradient of the negative log-likelihood obtained from running the particle filter with parameter $\B$ and observations $\y_{1:T}$.

    \subsubsection{Proximal update on the penalty term}
        \label{sec:prom_spar}
        Given that we can estimate the negative log-likelihood $\ell(\C; \y_{1:T})$ and its gradient $\nabla\ell(\C; \y_{1:T})$, we now need to account for the penalty term $\lambda R(\C)$.
        Note that the $L1$ penalty is not differentiable when a coordinate is $0$, hence it cannot be optimised using standard gradient descent if we aim to recover sparse estimates.
        It is however convex, and particularly well suited to the use of a proximity operator update \cite{combettes2011proximal}, that can be understood as an implicit subgradient step.

        The proximity operator update, for the $L1$ penalty, reads as a simple soft thresholding, so that our resulting scheme; combining both steps, is
            \begin{align}
                \begin{split}
                    \label{eq:soft_thresh}
                    \widetilde{\C}_s &= \mathrm{update}_\eta(\C_{s-1}, \nabla\ell(\C_{s-1}|\y_{1:T})),\\
                    {\C}_s &= \mathcal{T}_{\eta\cdot\lambda}(\widetilde{\C}_s),
                \end{split}
            \end{align}
        where the soft-thresholding operator, \[\mathcal{T}_\alpha(x) = \max(|x| - \alpha, 0)\cdot \mathrm{sgn}(x),\] is applied element-wise.
        The above algorithm belongs to the class of stochastic proximal gradient methods, the convergence of which is well studied, for instance in \cite{rosasco2020convergence, combettes2016stochastic}.

        Note that automatic differentiation can also be used when applying the $L1$ penalty, as the $L1$ penalty admits a subgradient \cite{jax2018github, paszke2019pytorch}. 
        However, the proximal operator update is more computationally efficient, and more stable. 
        The proximal operator is also more versatile, as it allows using other penalties such as low rank, or $L0$ penalty, which cannot be applied by the subgradient method \cite{bach2012optimization}.

    \subsection{S-GraphGrad algorithm}
    \label{sec:init_alg}
        We have now all the elements for solving \eqref{eq:objective_function}. 
        We first start with a full batch implementation, in Alg.~\ref{alg:graphgrad_nobatch}, which we call S-GraphGrad, and which operates on the entire series of observations at once.
        
            \begin{algorithm}[H]
            {
                \footnotesize
                \caption{Series GraphGrad algorithm  \hfill (S-GraphGrad) \hspace{1em}}
                \label{alg:graphgrad_nobatch}
                \begin{algorithmic}[1]
                    \Input Series of observations $\y_{1:T}$, number of steps $S$, penalty parameter $\lambda$, $\naturals^{N_x \times M}$ degree matrix $\D$, initial coefficient value $\C_0$, learning rate $\eta$.
                    \Output Sparse $\reals^{N_x \times M}$ matrix $\C$ of polynomial coefficients.
                    \For{$s = 1,\dots,S$}
                        \State Run Alg. \ref{alg:sgd_dpf} with $p(\x_{t}|\x_{t-1}; \C_{s-1}, \D)$ and observations $\y_{1:T}$. \label{step_graphgrad_nobatch:run_dpf}
                        \State Estimate $\ell(\C_{s-1}| \y_{1:T})$ and $\nabla \ell(\C_{s-1}| \y_{1:T})$ via Eq.~\eqref{eq:likelihood} and backpropagation. \label{step_graphgrad_nobatch:get_grads}
                        \State Set $\widetilde{\C}_{s} = \mathrm{update}_\eta(\C_{s-1}, \nabla \ell(\C_{s-1}| \y_{1:T}))$. \label{step_graphgrad_nobatch:grad_update}
                        \State Set ${\C}_s = \mathcal{T}_{\eta\cdot\lambda}(\widetilde{\C}_s)$. \label{step_graphgrad_nobatch:prox_update}
                    \EndFor
                    \State Output $\C = \C_S$.
                \end{algorithmic}
            }
            \end{algorithm}

    \subsubsection{S-GraphGrad description}

        S-GraphGrad takes as input the series of observations $\y_{1:T}$, the number of steps $S$, the penalty parameter $\lambda$, the $\naturals_0^{N_x \times M}$ degree matrix $\D$, the initial coefficient value $\C_0$, and the learning rate $\eta$, producing as output a sparse $\reals^{N_x \times M}$ matrix $\C$ of polynomial coefficients.
        S-GraphGrad iterates for $S$ steps, with the $s$-th step proceeding as follows.
        
        First, we run a differentiable particle filter with the estimate of the coefficients from the previous step $\C_{s-1}$ with observations $\y_{1:T}$ (line \ref{step_graphgrad_nobatch:run_dpf}).
        When running the filter, we assume that we either know or have suitable estimates of the observation model $p(\y_t|\x_t)$ and the proposal distribution $\pi(\x_t|\x_{t-1}, \y_{t})$, as well as the state noise. 
        For example, if the noise is additive and Gaussian, we have Eq.~\eqref{eq:poly_eval_early}, and would require a suitable estimate of $\bm\Sigma_v$.
        Other noises can be accounted for, such as multiplicative Gaussian, assuming a definition of how the estimated polynomial model and the noise interact to compute $p(\x_{t}|\x_{t-1}; \C, \D)$.

        While running the filter, we process the weights to obtain an estimate of the likelihood of the parameter $\C_{s-1}$ using Eq.~\eqref{eq:likelihood}, and use automatic differentiation to obtain an estimate of $\nabla \ell(\C^{s-1}|\y_{1:T})$ (line \ref{step_graphgrad_nobatch:get_grads}).
        We then apply a gradient-based update step, on the negative log-likelihood, of a given minimisation scheme $\mathrm{update}$ with learning rate $\eta$ to $\C_{s-1}$, yielding $\widetilde{\C}_{s}$.
        Namely, the gradients $\nabla \ell(\C_{s-1}|\y_{1:T})$ are those of the negative log-likelihood, as we are aiming to maximise the log-likelihood, and hence we minimise the negative log-likelihood using the minimisation scheme $\mathrm{update}$ (line \ref{step_graphgrad_nobatch:grad_update}).

        We then apply the proximal update at $\widetilde{\C}_{s}$, yielding ${\C}_s$, the estimated coefficient matrix at step $s$ (line \ref{step_graphgrad_nobatch:prox_update}).
        In the case of the $L1$ penalty the proximity operator is the soft thresholding operator, depending on the learning rate of $\eta$ and a penalty parameter of $\lambda$.

    \subsubsection{Combating likelihood degeneracy}
        \label{sec:setting_up}

        In practice, using S-GraphGrad is hampered by likelihood degeneracy, which causes the likelihood to evaluate numerically as zero due to the limited precision of floating point arithmetic.
        This could result in the gradient vanishing, and hence numerical failure of the scheme.
        
        A typical way to combat this is by using log weights, different floating point representations, or gradient scaling, but these do not address the core problem, which is that the likelihood becomes more concentrated the longer the observations series is.
        We will now discuss this phenomenon, and how we mitigate it.
        
        For observation series longer than a few elements, the likelihood function, given in Eq.~\eqref{eq:likelihood}, when evaluated with parameters that do not yield dynamics close to the true dynamics, is numerically $0$.
        Numerical zeros are a limitation of computer arithmetic, and in this case result from the multiplication of the very low likelihoods that occur in an unadapted filter to yield a result that is numerically zero.
        Note that, if operating on a log scale, we instead obtain $-\infty$ rather than 0, but the root cause is the same.
        Note that even when the dynamics are perfectly known we still observe this phenomenon, and here we aim to mitigate the effect of unadapted parameters, not the fundamental issue of likelihood accumulation in particle filters.

    {

        There are several standard ways to mitigate the effects of likelihood concentration within SSMs, such as variance inflation, where we increase the magnitude of the state and observation covariances to decrease likelihood concentration, or simply running the particle filter with a larger number of particles.
        However, both of these methods break down as the length of the observation series increases, as they do not address the underlying issue of the parameters not being adapted to the observed data.
        
        The likelihood degenerating causes the gradients of the log-likelihood to explode, and hence makes it impossible to perform gradient-based inference.
        We can address this issue by running our method multiple times, first to fit a coarse estimate, and then refining this estimate over several subsequent iterations.
        We fit this coarse estimate using only the first few observations, and then gradually introduce more observations, refining our estimate and mitigating likelihood degeneracy due to unadapted parameters.
    }

    \subsection{B-GraphGrad algorithm}
    \label{sec:graphgrad}
        We end this section by presenting B-GraphGrad, our final method for estimating the dynamics of a general non-linear SSM via a polynomial approximation.
        This method builds upon that described Alg.~\ref{alg:graphgrad_nobatch} (S-GraphGrad) in Sec.~\ref{sec:init_alg}, proposing a sequentially-batched implementation of the method, utilising an observation batching strategy to mitigate likelihood concentration issues.

        The proposed method is doubly iterative: we iterate over telescoping batches of observations, within which we iteratively estimate the coefficients of our polynomial approximation.
        We create $B$ batches of observations by $\y^{(b)} := \y_{1:\lceil bT/B \rceil}$ for $b = 1,\dots,B$.
        {Note that these batches are of increasing size, with $\y^{(b)} \subseteq \y^{(b+1)}$ for $b = 1,\dots,B$.}
        Within each batch of observations, we perform $S$ runs of Alg.~\ref{alg:graphgrad_nobatch} (S-GraphGrad). 
        We perform batching to avoid numerical errors, as the first sampled trajectories, with parameters close the the initial random initialisation, will likely have an extremely small log-likelihood, which compounds numerical errors when computing the weights in Alg.~\ref{alg:sgd_dpf} \cite{doucet2009tutorial}.
        Note that, in the case that the true system can represented as a polynomial, few batches may be needed.
        Indeed, in the case of the Lorenz 63 oscillator, we require only a batch of $10$ observations to initialise the coefficients, and can then proceed with estimation on series of lengths exceeding $1000$.
        However, in general the true system is not polynomial, so we proceed with fixed-size batches for simplicity and robustness.
        We present the method in Alg.~\ref{alg:graphgrad} (B-GraphGrad), and describe it below.

            \begin{algorithm}[H]
            {
                \footnotesize
                \caption{Batched GraphGrad algorithm \hfill (B-GraphGrad) \hspace{1em}}
                \label{alg:graphgrad}
                \begin{algorithmic}[1]
                    \Input Series of observations $\y_{1:T}$, number of batches $B$, steps per batch $S$, penalty parameter $\lambda$, learning rate $\eta$, maximal degree $d$, hidden state size $N_x$.
                    \Output Sparse $\reals^{N_x \times M}$ matrix $\C$ of polynomial coefficients.
                    \State Construct $\D$ by running Alg.~\ref{alg:generate_D}.
                    \State Randomly initialise $\C_{0} \in \reals^{N_x \times M}$ element-wise by sampling a $\mathcal{U}(-1,1)$ distribution.
                    \For{$b = 1,\dots,B$}
                        \State Set $\y^{(b)} := \y_{1:\lceil bT/B \rceil}$.
                        \State Run Alg. \ref{alg:graphgrad_nobatch} (S-GraphGrad) with observations $\y^{(b)}$, number of steps $S$, penalty parameter $\lambda$, degree matrix $\D$, initial coefficient value $\C_{b-1}$, and learning rate $\eta$, setting $\C_b$ to the output.
                    \EndFor
                    \State Output $\C := \C_B$
                \end{algorithmic}
            }
            \end{algorithm}

    \subsubsection{B-GraphGrad description}
        As stated in Sec.~\ref{sec:graphgrad}, B-GraphGrad builds upon S-GraphGrad, implementing observation batching to mitigate likelihood generation for unadapted models.
        In particular, B-GraphGrad runs for $B$ iterations, with the $b$-th iteration running S-GraphGrad with $S$ iterations on the observation series $\y^{(b)} := \y_{1:\lceil bT/B \rceil} \subseteq \y_{1:T}$.
        This allows us to `warm up' the coefficient parameter $\C$, so that when we are performing estimation on the entire series we do not encounter issues due to likelihood concentration. 
        We avoid these issues by learning $\C$ first on small series, which have relatively dispersed likelihoods due to their length.
        By learning an initial estimate of the transition dynamics on the initial batches, we have a better model when we come to estimating $\C$ on the longer series that can display likelihood issues when the model is not adapted.

        B-GraphGrad proceeds as follows.
        First, we construct the degree matrix $\D$ given the selected maximum degree $d$ and hidden state dimension $N_x$. 
        $\D$ is constructed following Alg.~\ref{alg:generate_D} described in Sec.~\ref{sec:gen_deg}.
        We note that $\D$ is a static parameter, and is not learnt through our training.
        We then generate an initial value for $\C$, denoted $\C_0$. 
        We can choose this value randomly, as we avoid likelihood related issues through our batching procedure.
        In Alg.~\ref{alg:graphgrad} (B-GraphGrad) we draw $\C_0$ element-wise by sampling a uniform $\mathcal{U}(-1,1)$ distribution, however in principle any real valued distribution could be used, such as a standard normal distribution.
        After initialising $\C$, we generate the $B$ observation batches.
        We then iterate over the $B$ batches of observations.
        For batch $b$, we run Alg.~\ref{alg:graphgrad_nobatch} (S-GraphGrad) with the parameter estimate from the previous batch, $\C_{b-1}$, and label the resulting updated parameter by $\C_b$.
        The $B$-th batch is the final batch and trains using the entire series of observations, outputting the final value $\C_B$, which is learnt so as to optimise our objective function Eq.~\eqref{eq:objective_function}.
        
    \section{Discussion}
        \label{sec:discussion}
        We now discuss our algorithm, B-GraphGrad, as given in Alg.~\ref{alg:graphgrad}, and provide some potential extensions and modifications to particular systems.

    \subsection{GraphGrad prerequisites}
        In order to apply GraphGrad, we must either know or have estimates of the observation model and its noise process, and must know the form of the state noise.
        The observation model are typically assumed known in dynamical systems, such as by the intrinsic properties of the sensors used, or are estimated from previous studies.
        We require that the likelihood of the observation is differentiable with respect to $\C$, as otherwise we cannot apply the differentiable particle filter.
        If we use a proposal distribution $\pi(\x_t|\x_{t-1}, \y_t) \neq p(\x_t|\x_{t-1}) = p(\x_t|\x_{t-1}; \C, \D)$, then we must be able to sample from this distribution differentiably with respect with $\C$, and it must admit a differentiable likelihood with respect with $\C$.
        
        Furthermore, we assume that the state noise is such that we can sample it in a differentiable manner.
        An example of such a noise is the Gaussian distribution, from which we can generate sample differentiably with regard to the distribution parameters using the re\cred{parameterisation} trick \cite{kingma2013auto}. 
        Many distributions can be sampled differentiably with regard to their parameters using similar tricks, such as the multivariate t distribution (with known degrees of freedom), which could be used if heavier tails are required.
        
    \subsection{Computational cost}
        B-GraphGrad, as given in Alg.~\ref{alg:graphgrad}, requires $SB$ evaluations of the particle filter to obtain the negative log-likelihood and its gradient.
        The particle filter is of time complexity $\mathcal{O}(KT),$ and therefore our method is of complexity $\mathcal{O}(SBKT)$. Indeed, we evaluate the particle filter $SB$ times, and obtain the gradients via reverse-mode automatic differentiation, which is of the same time complexity as the function we differentiate. 
        Note that we neglect here the complexity cost of sampling distributions and evaluating the likelihoods in the particle filter, as these vary from model to model, and occur the same number of times across filter runs.
        
        The cost of evaluating the polynomial in Eq.~\eqref{eq:polynomial} is small, and of complexity $\mathcal{O}(MN_x^2)$.
        We note that the polynomial evaluation can be efficiently parallelised as $\D$ is fixed, so it is possible to evaluate the polynomial by performing an associative scan over evaluating the monomials.
        That is, we can evaluate the $x_j^{D_{i,j}}$ terms in parallel, and then evaluate their product and sum in parallel.
        Furthermore, the vector-scalar element product is typically automatically abstracted to an SIMD operation, speeding evaluation up by a factor of $N_x$.

        Finally, we note that the computation of the particle filter and its gradient can be accelerated by observing that the computations in the differentiable particle filter depend on the previous state only through the particles and the weights.
        Under this restriction, we can implement the particle filter as a scan-with-carry.
        Therefore, it is possible to construct the computational graph of the particle filter, and therefore of its gradient, from the computational graph of the scanned function, yielding a much smaller graph than from the entire filter \cite{jax2018github}.

    \subsection{Exploiting parallelisation to decrease runtime}

        B-GraphGrad, if implemented following Alg.~\ref{alg:graphgrad}, is a sequential algorithm.
        Sequential operation is required as the estimate at each step depends on the estimate at the previous step.
        However, we can use parallel computing to reduce the elapsed time of the computation by computing fewer batches.

        In particular, the batching proposed in B-GraphGrad is very conservative, in that the batch size increases by the same increment in all instances.
        In practice, as the approximand uses few parameters, with each parameter having a distinct effect, we rapidly adapt to the system within the first few batches.

        We hence propose a more efficient procedure. 
        We initialise $P$ independent coefficient matrices, $\{\C^{(p)}\}_{p=1}^{P}$, and learn each independently in parallel for $B_I$ batches of observations.
        Then, we check if there exists a subset of the $P$ coefficient matrices such that the coefficient matrices are close in value, for example by attempting to find $\C \in \reals^{N_x \times M}$ such that  $||\C - \C^{(p)}|| < \epsilon \ \forall p \in \{1, \dots, P\}$ for some matrix norm $||\cdot||$ and $\epsilon > 0$. 
        If this matrix exists, then this indicates that the coefficient matrices have adapted to the system.
        We then cease batching, and learn on the entire series of observations using the coefficient matrices in our subset, aggregating after finishing optimisation, e.g., via an element-wise mean.
        If the subset cannot be constructed, we continue learning for another batch of observations, and then recheck the above condition, stopping when either we exhaust the set of observations or the subset of matrices can be constructed.
        
        This batching method can take advantage of parallel processing by performing optimisation on each independent parameter in parallel, with the subset construction and matrix calculations being cheap in comparison.
        This accelerates our inference by reducing the number of batches that need to be constructed, therefore reducing the run time of the algorithm.

    \subsection{\cblue{Interpretation as a library regression method}}
        \label{sec:discussion:library}
        \cblue{
        Our method, at its core, fits a series of polynomial functions aiming to recover the transition kernel of a SSM.
        Therefore, B-GraphGrad can be interpreted as performing function library regression, similar to SINdY \cite{brunton2016discovering}, with the function library comprising all polynomials in $N_x$ variables of degree less than or equal to $d$, and fitted performed simultaneously for each state dimension.
        }
        
        \cblue{
        Given this interpretation, it is possible to expand the library of terms to include additional terms, such as trigonometric or exponential terms.
        These terms would allow for an even larger class of systems to be exactly represented by our model, but come at the downside of requiring additional machinery to evaluate, whereas polynomial terms admit a simple vectorised expression in Eq.~\eqref{eq:polynomial}.
        However, non-polynomial terms should be evaluated as separate expressions, and then combined with the polynomial terms after computation.
        For example, if the system is defined in terms of angles, trigonometric terms could be included, or functions of the differences between states could be included for systems involving potentials.
        }

    \section{Numerical study}
        \label{sec:numerics}
        We here present our experimental results, to illustrate and discuss the performance of our proposed approach.
        We are interested in the recovery of the underlying model in dynamical systems described by polynomial ordinary differential equations (ODEs), namely the Lorenz 63 (Sec.~\ref{sec:exp:l63}) and Lorenz 96 (Sec.~\ref{sec:exp:l96}) oscillators, and estimating a non-polynomial system, the Kuramoto oscillator (Sec.~\ref{sec:exp:kura}). 
        In each case, we use an Euler discretisation to build a discretised form for the model, and we map it with our problem formulation where the goal is to recover an estimate of a ground truth matrix $\C$.
        
        We implement the proposed B-GraphGrad algorithm, \cblue{and compare it against the fitting of a polynomial of the same degree using a maximum likelihood scheme, which we denote pMLE, for polynomial maximum likelihood estimator.} 
        This scheme is identical to our B-GraphGrad scheme, except that we remove the proximal steps. 
        pMLE fits a fully dense model, so in its case we present only RMSE, as the sparsity metrics are predetermined (i.e., no sparsity is recovered).
 
    \subsection{Experimental setup}
        \label{sec:expsetup}
        For our numerical experiments, we use the following settings, unless stated otherwise.
        We use the Novograd optimisation scheme \cite{li2019jasper, ginsburg2019training}, for $\text{update}$ in Alg.~\ref{alg:graphgrad_nobatch} line \ref{step_graphgrad_nobatch:grad_update}, with a fixed learning rate of $\eta = 10^{-3}$.
        We split our observation series into $B$ batches such that $B = \lceil T/10 \rceil$, therefore giving a batch size increment of approximately $10$.
        We use $K=100$ particles in our particle filter in Alg.~\ref{alg:sgd_dpf}.
        $T$ denotes the length of the observation series.
        For a given polynomial degree $d$, we construct the degree matrix $\D$ following Sec.~\ref{sec:gen_deg}.
        
        Performance assessment is performed using several quantitative metrics, either based on the recovery of the sparse support of $\C$ (in terms of specificity, precision, recall, and F1 score), or of its entries (in terms of root mean square error, RMSE). 
        We define an element of $\C$ as numerically zero if it is lower than $10^{-6}$ in absolute value, as this is approximately the precision of single precision floating point arithmetic. 
        Perfect recovery of positive and negative values is indicated by $1.0$ in all sparsity metrics, and $0.0$ RMSE.
        
        We choose the penalty parameter $\lambda$ for a system with state dimension $N_x$ and maximal degree $d$ by tuning it on a synthetic system. 
        This system is not used to generate the data for fitting, and is only used to tune $\bm\lambda$.
        This system is such that it has the same $N_x$ dimension state and a maximal degree of $d$ as the model we are fitting.
        We generate the $\C$ matrix of this system such that it is $75\%$ sparse, with the dense elements drawn from $\mathrm{U}(-N_x, N_x)$, and then scaled such that the maximal singular value of $\C$ is $1$.
        We discretise this system using an Euler discretisation with a timestep $\Delta t$ equal to the timestep of the system we aim to estimate, adding zero mean Gaussian noise terms $\v$ and $\r$ to the state and observations respectively, with $\bm\Sigma_v = \bm\Sigma_r = \Delta t\Id_{N_x}$.
        We then optimise $\lambda$ via setting $\lambda = 10^{l}$, with $l$ chosen to maximise the accuracy of the estimated $\C$ of this system using B-GraphGrad, via $10$ iterations of bisection over the interval $[-5,2]$.

    \subsection{Lorenz 63 model}
        \label{sec:exp:l63}
        We start our experiments, with the Lorenz 63 system \cite{lorenz1963deterministic}, which we transform into an NLSSM using an Euler discretisation with a timestep of $\Delta t = 0.025$, 
        yielding the system
            \begin{equation}
                \setlength{\arraycolsep}{2.5pt}
                \thinmuskip=1mu
                \medmuskip=2mu plus 2mu minus 2mu
                \thickmuskip=3mu plus 5mu minus 2mu
                \begin{alignedat}{2}
                    \label{eq:l63_disc}
                        x_{1,t+1} &= x_{1,t} + \Delta t (\sigma(x_{2,t} - x_{1,t})) + \sqrt{\Delta t} v_{1,t+1},\\
                        x_{2,t+1} &= x_{2,t} + \Delta t (x_{1,t} (\rho - x_{3,t}) - x_{2,t}) +  \sqrt{\Delta t} v_{2,t+1},\\
                        x_{3,t+1} &= x_{3,t} + \Delta t (x_{1,t} x_{2,t} - \beta x_{3,t}) +  \sqrt{\Delta t} v_{3,t+1},
                \end{alignedat}
            \end{equation}
        with the observation model $p(\y_{t}|\x_{t}) = \mathcal{N}(\x_t, \bm\Sigma_r)$.
        Hence, $N_x = N_y = 3$.
        We choose $\x_0$ such that $\x_0$ is equal to $1$ in the first element, and $0$ elsewhere. 
        We set $\bm\Sigma_v = \bm\Sigma_r = \sigma^2 \Id_3$, with $\sigma = 1$ unless stated otherwise.
        Note that we present the true $\C$ and $\D$ matrices for this system in Sec.~\ref{sec:l63_example_section}
        We use $\rho = 28, \, \sigma = 10, \, \beta = 8/3$, as these parameters are known to result in a chaotic system, and we set $t_0 = 0$. 
        Our particle filter is initialised with $p(\x_0) = \mathcal{N}(\x_0, \Id_3)$.
        
        We average the results, for our method GraphGrad, and pMLE, \cblue{a scheme that fits the same coefficient matrix $\C$ but does not promote sparsity}, on $150$ independent realisations of the specified dynamical system. 
        Hereafter we present our results, on three scenarios, namely assuming maximum degree $d=2$ or $3$, and varying the series length, then considering a varying noise amplitude.

    \subsubsection{Varying series length, maximum degree $d=2$}
    \label{sec:exp:l63d2_t}
    
            \begin{table}[ht]
            	\centering
            	\footnotesize
                \captionsetup{justification=centering, labelsep=newline}
                \caption{Lorenz 63: Average recovery metrics for variable series length for 150 independent systems. Maximum polynomial degree $d = 2$. }
                \setlength{\tabcolsep}{3pt}
            	
            	\label{tab:variablesystems_d2}
                	\begin{tabular}{|r|c||c|c|c|c|c|}
                		\hline
                		method & $T$ & RMSE ($10^{-3}$) & spec. & recall & prec. & F1\\
                		\hline
                		\hline
                        B-GraphGrad & $25$ & 1.6 & 0.90 & 0.92 & 0.96 & 0.94\\
                        pMLE & $25$ & 2.4 & - & - & - & -\\
                		\hline
                		  B-GraphGrad & $50$ & 1.0 & 0.98 & 0.97 & 0.98 & 0.98\\
                		pMLE & $50$ & 1.6 & - & - & - & -\\
                		\hline
                        B-GraphGrad & $100$ & 0.4 & 1.00 & 1.00 & 1.00 & 1.00\\
                        pMLE & $100$ & 0.8 & - & - & - & -\\
                		\hline
                        B-GraphGrad & $200$ & 0.2 & 1.00 & 1.00 & 1.00 & 1.00\\
                        pMLE & $200$ & 0.3 & - & - & - & -\\
                		\hline
                	\end{tabular}
            \end{table}

            \begin{figure}[ht]
                \centering
                \input{plots_tex/L63_varT_d2}
                \input{plots_tex/plot_leg_L96}
                \caption{Comparison of B-GraphGrad with pMLE over variable series length on the Lorenz 63 oscillator, with maximum polynomial degree $d=2$. Markers denote mean performance, with the ribbons being symmetric 95\% intervals.}
                \label{fig:serieslength_L63_d2}
            \end{figure}

            \begin{table}[ht]
            	\centering
            	\footnotesize
                \captionsetup{justification=centering, labelsep=newline}
                \caption{Lorenz 63: Median runtime relative to $T=25$ for variable series length for 150 independent systems. Maximum polynomial degree $d = 2$. }
                \setlength{\tabcolsep}{3pt}
            	
            	\label{tab:variablesystems_d2_runtime}
                	\begin{tabular}{|r|c||c|}
                		\hline
                		method & $T$ & Relative runtime\\
                		\hline
                		\hline
                        B-GraphGrad & $25$ & 1\\
                		\hline
                		  B-GraphGrad & $50$ & 3.86\\
                		\hline
                        B-GraphGrad & $100$ & 15.64\\
                		\hline
                        B-GraphGrad & $200$ & 62.25\\
                		\hline
                	\end{tabular}
            \end{table}
        Table~\ref{tab:variablesystems_d2} presents the results for learning over a range of series lengths $T$ with a maximal degree of $d=2$.
        In this case, estimating $\C$ requires learning $30$ parameters.
        Here, we assumed the knowledge of the correct degree of the underlying system, so these results serve to provide a baseline for the performance of our method in the scenario where the degree is known.
        We note that, for many dynamical systems, a default degree of $2$ is a sensible modelling choice, as this type of interaction in the rate is very common in many systems, such as in chemical and biological networks, population modelling, and meteorological systems.
        We observe that the performance of our method improves as the number of observations $T$ increases, indicating that our method well incorporates information from new observations.
        Furthermore, we see that, even for a small number of observations, our method recovers the system connectivity well, even if the RMSE is not as good for low values of $T$.
        This is due to changes in the system connectivity and the presence or absence of system terms having a large effect, and thus being relatively easy to infer compared the the specific value of the terms, the effects of which are obscured by the noise inherent to the system.
        Overall, we observe excellent performance, with good recovery of all parameters at all tested series lengths.
        We see that pMLE does not perform well, mostly because it recovers a fully dense system, and therefore yields matrix $\C$ with many \cred{nonzero} parameters that are, in truth, zero.
        Hence, the RMSE is poor, as the value of the truly \cred{nonzero} parameters is affected by the value of the falsely \cred{nonzero} parameters.

        From the results given in Table~\ref{tab:variablesystems_d2_runtime}, we see that our runtime seems to scale quadratically in $T$. 
        This follows from the linear complexity of the particle filter in $T$, and the number of batches $B$ also scaling linearly in $T$ as per the experimental setup.
        Each batch runs a particle filter of a fixed length, with each batch running a longer filter than the previous, and therefore the algorithm runtime scales quadratically with $T$.
        However, one can use the batching method presented in the introduction of Sec.~\ref{sec:graphgrad} to remove the quadratic scaling by fixing the number of batches, in which case the runtime scales linearly in $T$.

    \subsubsection{Varying series length and maximum degree $d=3$}
        
            \begin{table}[ht]
            	\centering
            	\footnotesize
                \captionsetup{justification=centering, labelsep=newline}
                \caption{Lorenz 63 average recovery metrics for variable series length for 150 independent systems. Maximum allowed degree $d = 3$.}
                \setlength{\tabcolsep}{3pt}
            	
            	\label{tab:variablesystems_d3}
                	\begin{tabular}{|r|c||c|c|c|c|c|}
                		\hline
                		method & $T$ & RMSE ($10^{-3}$) & spec. & recall & prec. & F1\\
                		\hline
                		\hline
                        B-GraphGrad & $25$ & 2.0 & 0.84 & 0.90 & 0.92 & 0.91\\
                        pMLE & $25$ & 2.8 & - & - & - & -\\
                		\hline
                		B-GraphGrad & $50$ & 1.5 & 0.96 & 0.95 & 0.97 & 0.96\\
                		pMLE & $50$ & 2.0 & - & - & - & -\\
                		\hline
                        B-GraphGrad & $100$ & 0.7 & 0.97 & 0.98 & 0.97 & 0.97\\
                        pMLE & $100$ & 1.3 & - & - & - & -\\
                		\hline
                        B-GraphGrad & $200$ & 0.4 & 1.00 & 1.00 & 1.00 & 1.00\\
                        pMLE & $200$ & 1.0 & - & - & - & -\\
                		\hline
                	\end{tabular}
            \end{table}

            \begin{figure}[ht]
                \centering
                \input{plots_tex/L63_varT_d3}
                \input{plots_tex/plot_leg_L96}
                \caption{Comparison of B-GraphGrad with pMLE over variable series length on the Lorenz 63 oscillator, with maximum polynomial degree $d=3$. Markers denote mean performance, with the ribbons being symmetric 95\% intervals.}
                \label{fig:serieslength_L63_d3}
            \end{figure}
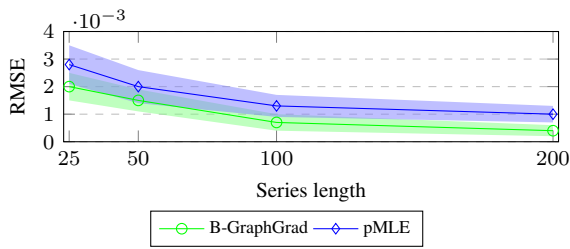
        Table~\ref{tab:variablesystems_d3} presents the results of our GraphGrad method over a range of series lengths $T$ with maximal degree of $3$.
        In this case estimating $\C$ requires estimating $60$ parameters.
        A degree of $3$ is greater than the true degree of the system, which, as we recall, is not in general known beforehand.
        Systems with a degree of $3$ are uncommon, however it is possible to approximate non-polynomial systems, with higher degree allowing better approximation of the dynamics.
        For example, the Kuramoto oscillator \cite{kuramoto1984chemical} can be well approximated via a centralised Taylor approximation, which our method can learn directly.
        
        We observe that the performance of our method improves as the number of observations increases, indicating that our method well incorporates information from new observations, and does not over fit given the over specification of the model relative to the system we are learning.
        We note a decrease in performance relative to Table~\ref{tab:variablesystems_d2} where the degree is that of the underlying system, however the results improve as $T$ increases.
        
        Furthermore, we note that the change in RMSE is more severe than the changes in sparsity metrics, as a small number of degree 3 terms are fit, where in reality they should be zero.
        As RMSE is computed only on terms recovered as non-zero, these contribution of these deviations is large relative to that of the degree 2 terms.
        We see that our method performs well, with the sparsity metrics being close to those of the $d=2$ system.
        The RMSE is inferior to the $d=2$ system, but still significantly outperforms the comparable polynomial MLE.

    \subsubsection{Variable noise magnitude}
    
            \begin{table}[ht]
            	\centering
            	\footnotesize
                \captionsetup{justification=centering, labelsep=newline}
                \caption{Lorenz 63: Average recovery metrics for variable noise magnitude for 150 independent systems. Maximum polynomial degree $d = 2$. $T=50$.}
                \setlength{\tabcolsep}{3pt}
            	
            	\label{tab:variablesystems_sigma}
                	\begin{tabular}{|r|c||c|c|c|c|c|}
                		\hline
                		method & $\sigma^2$ & RMSE ($10^{-3}$) & spec. & recall & prec. & F1\\
                		\hline
                		\hline
                        B-GraphGrad &$0.01$ & 0.09 & 1.00 & 1.00 & 1.00 & 1.00\\
                		pMLE & $0.01$ & 0.3 & - & - & - & -\\
                		\hline
                		B-GraphGrad &$0.1$ & 0.3 & 1.00 & 1.00 & 1.00 & 1.00\\
                		pMLE & $0.1$ & 0.6 & - & - & - & -\\
                		\hline
                        B-GraphGrad & $1$ & 1.0 & 0.98 & 0.97 & 0.98 & 0.98\\
                		pMLE & $1$ & 1.6 & - & - & - & -\\
                		\hline
                        B-GraphGrad & $5$ & 1.8 & 0.90 & 0.85 & 0.87 & 0.86\\
                		pMLE & $5$ & 2.3 & - & - & - & -\\
                		\hline
                	\end{tabular}
            \end{table}

            \begin{figure}[ht]
                \centering
                \input{plots_tex/L36_varsigma2_d2}
                \input{plots_tex/plot_leg_L96}
                \caption{Comparison of B-GraphGrad with pMLE over variable noise magnitude on the Lorenz 63 oscillator, with maximum polynomial degree $d=2$. Markers denote mean performance, with the ribbons being symmetric 95\% intervals.}
                \label{fig:serieslength_L63_s2}
            \end{figure}
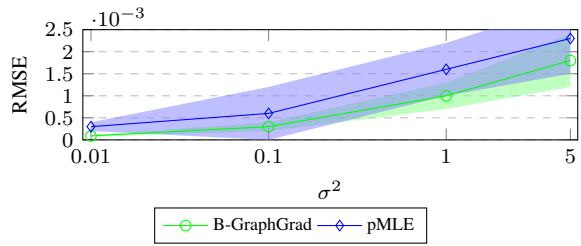

        In Table~\ref{tab:variablesystems_sigma}, we present the results of our method over a range of noise magnitudes $\sigma^2$, now fixing the number of observations to $T=50$.
        We remind that the results from Tables~\ref{tab:variablesystems_d2} and \ref{tab:variablesystems_d3}, were obtained using $\sigma^2 = 1$.
        We observe that our method performs well for all tested values of $\sigma^2$, especially in sparsity metrics.
        As expected, the recovery quality degrades as the signal to noise ratio decreases when we increase $\sigma^2$.
        We note that a large $\sigma^2$ affects both the system and the observation, so increasing it has a particularly pronounced effect.

    \subsubsection{Likelihood degeneracy}

        In Sec.~\ref{sec:setting_up}, we discussed the effect of likelihood concentration on parameter estimation in the particle filter.
        We demonstrate this phenomenon below, giving the mean number of timesteps before the likelihood becomes numerically zero for a stochastic variant of the Lorenz 63 model, given by Eq.~\eqref{eq:l63_disc_section}.

        \begin{table}[ht]
            \centering
            \footnotesize
            \captionsetup{justification=centering, labelsep=newline}
            \caption{Average number of timesteps/observations ($\Delta t = 0.025$) before likelihood degeneracy with a random $\C$ matrix for the Lorenz 63 model ($d = 2$) over 200 independent systems initialised at random points.}
            \setlength{\tabcolsep}{2pt}
            
            \label{tab:l63_degeneracy}
                \begin{tabular}{|c||c|c|c|c|c|c|c|c|}
                    \hline
                    number of particles $K$ & 5 & 10 & 20 & 50 & 100 & 250 & 500 & 1000 \\
                    \hline
                    timesteps before degeneracy & 7.8 & 12.2 & 15.7 & 23.8 & 35.3 & 57.5 & 76.8 & 111.3\\
                    \hline
                \end{tabular}
        \end{table}

         It is clear from Table \ref{tab:l63_degeneracy} that increasing the number of particles does combat the likelihood degeneracy, though it does not solve it, with computational cost increasing significantly for small gains.
         It is for this reason that we do not display the results of \cred{GraphGrad}, instead displaying only B-GraphGrad, as \cred{GraphGrad} fails to converge using single precision arithmetic due to these likelihood issues.

    \cblue{\subsubsection{Varying maximal degree $d$}}
        \cblue{The required number of parameters to estimate increases as $d$ increases, and therefore the computational cost increases as well. 
        In the Lorenz 63 oscillator we have $N_x = 3$, and therefore for $d=2$, the true degree of the system, we have to estimate $30$ parameters; $10$ parameters per state dimension.
        For $d=3$ this increases to $60$ parameters, which is $20$ parameters per state dimension.
        We present in Table~\ref{tab:l63_vary_d} the cost of running B-GraphGrad with different $d$ parameters relative to the cost of running with $d=2$. 
        All parameters are set per Sec.~\ref{sec:exp:l63d2_t}, except $d$, which we vary.
        We present the results in Table~\ref{tab:l63_vary_d}.
        }

            \begin{table}[ht]
            	\centering
            	\footnotesize
                \captionsetup{justification=centering, labelsep=newline}
                \caption{Lorenz 63: Average relative runtime of B-GraphGrad when run with different values of $d$.
                Runtime is presented relative to $d=2$.
                }
                \setlength{\tabcolsep}{3pt}
            	
            	\label{tab:l63_vary_d}
                	\begin{tabular}{|c||c|c|c|c|c|c|}
                        \hline
                        maximum polynomial degree $d$ & 1 & 2 & 3 & 4 & 5 & 6 \\
                        \hline
                        average relative runtime & 0.41 & 1.0 & 1.98 & 3.46 & 5.52 & 8.27 \\
                        \hline
                        relative number of parameters & 0.4 & 1.0 & 2.0 & 3.5 & 5.6 & 8.4 \\
                        \hline
                    \end{tabular}
            \end{table}
        \cblue{
        From Table~\ref{tab:l63_vary_d}, we observe that the runtime of our method scales as the number of rows in $\C$, or equivalently, the runtime grows as the number of elements of $\C$ increase.
        For example, there are $84$ monomials in $3$ variables of of degree $d \leq 6$, and we observe that running our method with $d=6$ takes $8.27$ times longer than with $d=2$ (which has $10$ monomials).
        The runtime foe larger values of $d$ is slightly lower than expected, due to constant computational overhead introduced in other areas of the implementation.
        This overhead makes up proportionally more of the runtime for smaller values of $d$, therefore resulting in smaller relative runtimes for larger values of $d$.
        Finally, we note that this experiment was performed without parallelising the evaluation of Eq.~\eqref{eq:polynomial} or its gradients.
        When parallelisation is enabled, the relative runtime is limited by computational resources, and in the presence of unlimited resources the runtimes are approximately equal.
        However, the total number of operations performed will scale similarly to as before.
        }
    \cred{F}

    \subsubsection{Prox update compared to subgradient update}
        B-GraphGrad uses the proximal operator of the $L1$ norm to apply a sparsity promoting penalty, following Sec.~\ref{sec:prom_spar}.
        However, as noted, we can also use subgradient methods to minimise our cost function including the $L1$ penalty.
        Using standard convex analysis definition, a subgradient of the $L1$ norm at an element $x$ (i.e., an element of the subdifferential of L1 function, at $x$), is a vector of same dimension than $x$, with entries equal to $\mathrm{sgn}(x)$, using the convention $\text{sgn}(0) = 0$. Note that, numerically, we never encountered exactly zero entries when running the subgradient descent method, and that the subgradient at a $0^-$ (resp. $0^+$) entry is taken as $-1$ (resp. $+1$).
        We use the same learning rate of $\eta = 10^{-3}$ than in the proximal version, and the same $(B,S)$ parameters for the number of inner/outer iterations.
        The computational cost differences within each step are negligible, however the proximal method is overall more efficient, as the proximal version appears to converge faster.

        We test both the subgradient approach and the proximal approach on the same system as Sec.~\ref{sec:exp:l63d2_t}. 
        We present both the average absolute value of elements where the ground truth is $0$, and the RMSE of the estimate.
        Parameters are set per Sec.~\ref{sec:exp:l63d2_t}.

            \begin{table}[ht]
            	\centering
            	\footnotesize
                \captionsetup{justification=centering, labelsep=newline}
                \caption{Lorenz 63: Average absolute value of elements of $\C$ where ground truth is $0$ over 150 independent systems. Maximum polynomial degree $d = 2$.}
                \setlength{\tabcolsep}{3pt}
            	
            	\label{tab:prox_vs_subgradient}
                	\begin{tabular}{|r|c||c|c|c|c|c|}
                		\hline
                		method & $T$ & Recall & RMSE ($10^{-3}$)\\
                		\hline
                		\hline
                        B-GraphGrad (prox) & $25$ & $0.92$ & $1.6$\\
                        B-GraphGrad (subgrad) & $25$ & $0.76$ & $2.0$\\
                		\hline
                        B-GraphGrad (prox) & $50$ & $0.97$ & $1.0$\\
                        B-GraphGrad (subgrad) & $50$ & $0.80$ & $1.4$\\
                		\hline
                        B-GraphGrad (prox) & $100$ & $1.00$ & $0.4$\\
                        B-GraphGrad (subgrad) & $100$ & $0.91$ & $0.6$\\
                		\hline
                        B-GraphGrad (prox) & $200$ & $1.00$ & $0.2$\\
                        B-GraphGrad (subgrad) & $200$ & $0.93$ & $0.4$\\
                		\hline
                	\end{tabular}
            \end{table}

        The proximal operator method is more efficient per iteration, in particular, we obtain estimates closer to the ground truth in the same number of iterations as the subgradient method.
        Further, as the methods have negligable difference in computational cost, this illustrates the superiority of the proximal over the subgradient method for this problem.

    \subsection{Lorenz 96 model}
        \label{sec:exp:l96}

\begin{figure}[ht]
            \centering
            \includegraphics[width=0.45\linewidth]{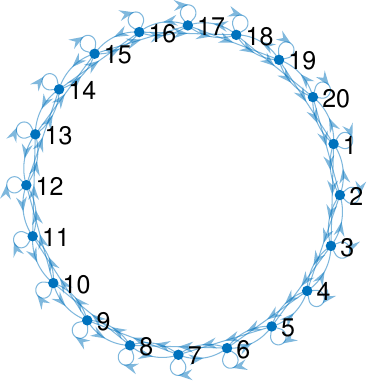}
            \includegraphics[width=0.45\linewidth]{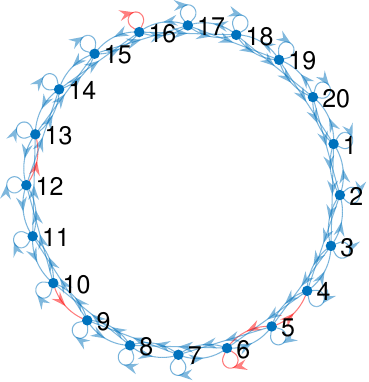}
            \caption{Graph encoding the connectivity of the Lorenz 96 system for $N_x = 20$.
            The left plot represents the true connectivity of the system under the graphical perspective described in Section \ref{sec:graphical_matrices}.
            The right plot represents the connectivity retrieved by the B-GraphGrad algorithm.
            The links in blue correspond to connections where the monomial is also correctly identified.
            The links in red correspond to connections where the monomial is incorrectly identified. 
            In this example, the node connectivity was perfectly retrieved by B-GraphGrad, however in 6 out of the 80 links, the link was found but with an incorrect monomial (e.g., the self loop of node 16 was identified as quadratic while it should be a linear term).}
            \label{fig:l96_graph}
        \end{figure}

        The Lorenz 63 system, while well studied and challenging to estimate, is only 3 dimensional.
        In order to test the applicability of our method to higher dimensional chaotic systems, we test on the Lorenz 96 system \cite{lorenz1996predictability},
        which we transform into a NLSSM using an Euler discretisation, yielding the system 
            \begin{equation}
                \begin{alignedat}{2}
                    d_{i,t+1} &= x_{i-1, t}(x_{i+1, t} - x_{i-2, t}) - x_{i, t} + F, \\
                    x_{i, t+1} &= x_{i, t} + \Delta t \cdot d_{i,t+1} +  \sqrt{\Delta t} \cdot v_{i,t+1}, \\
                    y_{i, t+1} &= x_{i, t+1} +  \sqrt{\Delta t} \cdot  r_{i, t+1},
                \end{alignedat}
            \end{equation}
        for $i = \{1, \dots, N_x\}, $ with $\v_{t} \sim \mathcal{N}(\bm0, \bm\Sigma_v), \r_{t} \sim \mathcal{N}(\bm0, \bm\Sigma_r)$, where $x_{, \cdot} = x_{N_x}$, $x_{-1, \cdot} = x_{N_x-1}$, and $x_{-2, \cdot} = x_{N_x-2}$. 
        For this experiment we choose $N_x = N_y = 20$.
        Therefore, in the case of $d=2$, estimating $\C$ requires estimating $4620$ parameters, and the case of $d=3$ estimating $\C$ requires estimating $35420$ parameters.
        We choose a forcing constant of $F = 8$, which is known known to result in a chaotic system.
        The connectivity graph of this system, constructed following Sec.~\ref{sec:graphical_matrices}, is given in Figure~\ref{fig:l96_graph}, where we observe that state $i$ is affected by states $i-2, i-1, i,$ and $i+1$, with index boundaries such that state $N_x \cdot k + i$ is equivalent to state $i$ for any $k \in \integers$.
        
        We set $\bm\Sigma_v = \bm\Sigma_r = \sigma^2 \Id_{N_x}$, with $\sigma = 1$ unless stated otherwise.
        We discretise this system with a timestep of $\Delta t = 0.025$.
        We choose $t_0 = 0$, and fix $\x_0$ such that $\x_0$ is equal to $1$ in the first element, and $0$ elsewhere.
        Our particle filter is initialised with $p(\x_0) = \mathcal{N}(\x_0, \Id_{N_x})$.
        We note that the system is of polynomial form, and therefore can be exactly recovered by our model, assuming $d$ equal or larger than two.

            \begin{table}[ht]
            	\centering
            	\footnotesize
                \captionsetup{justification=centering, labelsep=newline}
                \caption{Lorenz 96: Average recovery metrics for variable series length for 150 independent systems. Maximum polynomial degree $d = 2$. }
                \setlength{\tabcolsep}{3pt}
            	
            	\label{tab:variablesystems_d2_l96}
                	\begin{tabular}{|r|c||c|c|c|c|c|}
                		\hline
                		method & $T$ & RMSE ($10^{-3}$) & spec. & recall & prec. & F1\\
                		\hline
                		\hline
                        B-GraphGrad & $25$ & 1.8 & 0.92 & 0.93 & 0.93 & 0.93\\
                        pMLE & $25$ & 2.6 & - & - & - & -\\
                		\hline
                		B-GraphGrad & $50$ & 1.4 & 0.96 & 0.95 & 0.94 & 0.95\\
                        pMLE & $50$ & 2.0 & - & - & - & -\\
                		\hline
                        B-GraphGrad & $100$ & 0.8 & 0.99 & 0.97 & 0.98 & 0.98\\
                        pMLE & $100$ & 1.3 & - & - & - & -\\
                		\hline
                        B-GraphGrad & $200$ & 0.3 & 1.00 & 1.00 & 1.00 & 1.00\\
                        pMLE & $200$ & 0.8 & - & - & - & -\\
                		\hline
                	\end{tabular}
            \end{table}

            \begin{figure}[ht]
                \centering
                \input{plots_tex/L96_varT_d2}
                \input{plots_tex/plot_leg_L96}
                \caption{Comparison of B-GraphGrad with pMLE over variable series length on the Lorenz 96 oscillator, with maximum polynomial degree $d=2$. Markers denote mean performance, with the ribbons being symmetric 95\% intervals.}
                \label{fig:serieslength_L96_d2}
            \end{figure}
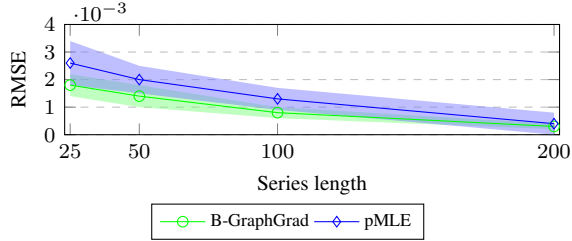

            \begin{table}[ht]
            	\centering
            	\footnotesize
                \captionsetup{justification=centering, labelsep=newline}
                \caption{Lorenz 96: Average recovery metrics for variable series length for 150 independent systems. Maximum polynomial degree $d = 3$. }
                \setlength{\tabcolsep}{3pt}
            	
            	\label{tab:variablesystems_d3_l96}
                	\begin{tabular}{|r|c||c|c|c|c|c|}
                		\hline
                		method & $T$ & RMSE ($10^{-3}$) & spec. & recall & prec. & F1\\
                		\hline
                		\hline
                        B-GraphGrad & $25$ & 2.4 & 0.80 & 0.92 & 0.73 & 0.82\\
                        pMLE & $25$ & 0.32 & - & - & - & -\\
                		\hline
                		B-GraphGrad & $50$ & 2.0 & 0.89 & 0.96 & 0.86 & 0.91\\
                        pMLE & $50$ & 2.5 & - & - & - & -\\
                		\hline
                        B-GraphGrad & $100$ & 1.1 & 0.95 & 0.98 & 0.96 & 0.97\\
                        pMLE & $100$ & 2.2 & - & - & - & -\\
                		\hline
                        B-GraphGrad & $200$ & 0.5 & 1.00 & 1.00 & 1.00 & 1.00\\
                        pMLE & $200$ & 1.6 & - & - & - & -\\
                		\hline
                	\end{tabular}
            \end{table}

            \begin{figure}[ht]
                \centering
                \input{plots_tex/L96_varT_d3}
                \input{plots_tex/plot_leg_L96}
                \caption{Comparison of B-GraphGrad with pMLE over variable series length on the Lorenz 96 oscillator, with maximum polynomial degree $d=3$. Markers denote mean performance, with the ribbons being symmetric 95\% intervals.}
                \label{fig:serieslength_L96_d3}
            \end{figure}
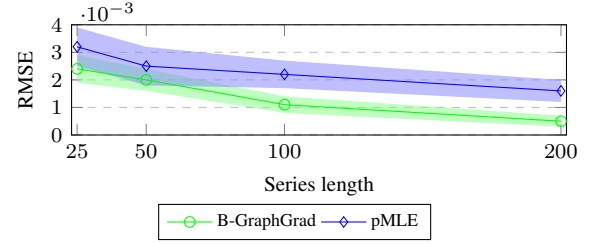

        We see that in both cases of $d=2$ and $d=3$, GraphGrad performs well, outperforming pMLE by a considerable margin.
        We see no significant deterioration of performance in GraphGrad in this setting, noting that we are estimating $4620$ parameters in the $d=2$ case, and $35420$ parameters in the $d=3$ case.

    \subsection{Kuramoto oscillator}
        \label{sec:exp:kura}
    
        The Kuramoto oscillator \cite{kuramoto1984chemical} is a mathematical model that describes the behaviour of a system of phase-coupled oscillators. 
        The model is described, for all $i \in \{1, \dots, N_x\}$, by
            \begin{align}
                \begin{split}
                \label{eq:kuramoto_eq}
                    \frac{\mathrm{d}\phi_i}{\mathrm{d}t} = \eta_i + N_x^{-1}\sum_{o=1}^{N_x}K\sin(\phi_i - \phi_{o}),
                \end{split}
            \end{align}
        where $\phi_i \in \reals$ denotes the phase of the $i$-th oscillator, and $K \in \reals$ is the coupling constant between oscillators.
        However, this does not restrict $\bm\phi$, which will, in general, diverge to $\pm \infty$ as $t \rightarrow \infty$.
        To address this, we transform Eq.~\eqref{eq:kuramoto_eq} by introducing derived parameters $R$ and $\psi$ such that
            \begin{align}
                \begin{split}
                \label{eq:kuramoto_eq_imag_reparam}
                    R\exp(\sqrt{-1}\psi) &= N_x^{-1} \sum_{o=1}^{N_x} \exp(\sqrt{-1}\phi_{o}),\\
                    (\forall i \in \{1,\ldots,N_x\})\; \frac{\mathrm{d}\phi_i}{\mathrm{d}t} &= \eta_i + KR \sin(\psi - \phi_i),
                \end{split}
            \end{align}
        which restricts $\bm\phi \in [-\pi, \pi]^{N_x}$.
        This is done purely for computational reasons, and does not affect the properties of the system or interpretation of $\bm\phi$ in any way.
        We transform Eq.~\eqref{eq:kuramoto_eq_imag_reparam} into a NLSSM using an Euler discretisation, yielding
            \begin{equation}
                \label{eq:kuramoto_nlssm}
                \begin{alignedat}{2}
                    R\exp(\sqrt{-1}\psi) &= N_x^{-1} \sum_{o=1}^{N_x} \exp(\sqrt{-1}x_{o, t}),\\
                    d_{i, t+1} &= \eta_i + KR\sin(\psi - x_i),\\
                    x_{i, t+1} &= x_{i, t} + \Delta t d_{i, t+1} +  \sqrt{\Delta t} v_{i, t+1}, \\
                    y_{i, t+1} &= x_{i, t+1} +  \sqrt{\Delta t} r_{i, t+1},
                \end{alignedat}
            \end{equation}
        for $i \in \{1, \dots, N_x\}, $where $\x$ denotes the phases in the discretised system to ease comparison with the rest of the work.
        We choose $N_x = 20$, and $K = 0.8$.
        We set $\bm\Sigma_v = \bm\Sigma_r = \sigma^2 \Id_{N_x}$, with $\sigma = 0.1$. 
        We discretise this model with a timestep of $\Delta t = 0.05$.
        We sample $\eta_i \sim \mathcal{N}(0.5, 0.5^2)$ and $\x_{i,0} \sim U(-\pi, \pi) \ \forall i \in \{1, \dots, N_x\}$.
        Our particle filter is initialised with $p(\x_0) = \mathcal{N}(\x_0, 0.2\Id_{20})$.
        We run the system until $t=10$, and then begin collecting observations.
        
        We note that this model cannot be represented exactly as a polynomial, and therefore the classification metrics used to illustrate sparsity recovery do not make sense.
        Therefore, we only present a normalised RMSE metric, where we compare the relative error in recovering the hidden state.
        nRMSE is given by $\mathrm{nRMSE}(\x_{EST}, \x_{TM}, \x_{GT}) = \frac{\mathrm{RMSE}(\cred{\x_{GG}}, \x_{GT})}{\mathrm{RMSE}(\x_{TM}, \x_{GT})}$, where $\x_{EST}$ is the sequence of means recovered by an estimator, $\x_{TM}$ is the sequence of means recovered by a particle filter running with the true model, and $\x_{GT}$ is the sequence of states we generate when creating our synthetic data.
        Therefore, $\mathrm{nRMSE} = 1$ indicates identical performance in terms of state recovery between the true model and an approximation.

        We compare the performance of GraphGrad against the pMLE as above, but also against the true model, where we estimate the parameters using maximum likelihood, which we denote by TrueMLE.
        TrueMLE is an oracle algorithm, and serves as a baseline for the case there the form of the model is known.
        TrueMLE assumes the form the model in Eq.~\eqref{eq:kuramoto_nlssm} is known, and estimates the $\bm\eta$ and $K$ parameters using maximum likelihood.

        \begin{figure}[ht]
            \centering
            \input{plots_tex/kura_varT}
            \input{plots_tex/plot_leg_kura}
            \caption{Comparison of B-GraphGrad with pMLE and TrueMLE over variable series length on the Kuramoto oscillator. Quantities are divided by the RMSE of a filter utilising the true model. Markers denote mean performance, with the ribbons being symmetric 95\% intervals.}
            \label{fig:serieslength_kura}
        \end{figure}
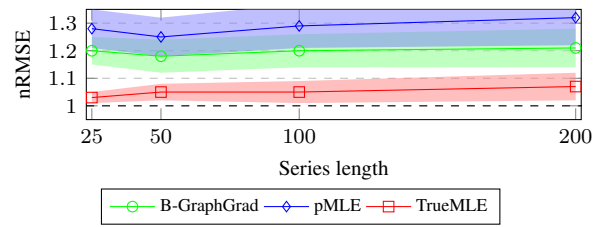

        We see in Figure~\ref{fig:serieslength_kura} that TrueMLE gives the best results, which is expected given that it is an oracle algorithm with respect to the form of the dynamics, whilst B-GraphGrad estimates both the form of the dynamics and the parameters thereof.
        Within these methods, GraphGrad performs significantly better than the polynomial MLE, and on average has only $20\%$ greater RMSE than state recovery with the true model and known parameters, whilst also assuming no knowledge of the underlying dynamics.

    \section{Conclusion}
        \label{sec:conclusion}
        In this work, we have proposed GraphGrad, a method for fitting a sparse polynomial approximation to the transition function of a general SSM.
        GraphGrad utilises a differentiable particle filter to learn a polynomial \cred{parameterisation} of a general SSM using gradient methods, from which we infer the connectivity of the state.
        GraphGrad promotes sparsity using a proximal operator, which is computationally efficient and stable.
        Our method is memory efficient, requiring only to keep track of a few parameters.
        Moreover, it is expressive, as many well known systems can be exactly represented by polynomial rates.
        Our method can utilise long observation series without vanishing gradients as we utilise observation batching to mitigate likelihood degeneracy.
        The method displays strong performance inferring the connectivity of a chaotic system, and keeps performing well when the true system cannot be exactly represented by the approximation.
    {
    \balance
    \bibliographystyle{IEEEtran}
    \bibliography{IEEEabrv,citeme.bib}
    }
\end{document}

%% file: plots_tex/L63_varT_d2.tex
\begin{tikzpicture}
\begin{axis}[
    xlabel={\footnotesize Series length},
    ylabel={\footnotesize RMSE},
    xmin=23.0, xmax=202.0,
    ymin=0.0, ymax=0.004,
    xtick={25, 50, 100, 200},
    ytick={0.000, 0.001, 0.002, 0.003, 0.004},
    legend style={at={(0.5,-0.3)},anchor=north,font=\footnotesize},
    ymajorgrids=true,
    grid style=dashed,
    ticklabel style = {font = \footnotesize},
    yticklabel style={/pgf/number format/fixed},
    height = 1.2in,
    width = 0.45\textwidth,
]


\addplot [name path=upper_gg_L63_d2,draw=none] table[x=x,y expr=\thisrow{y}+\thisrow{err}] {gg_L63_t_d2.dat};
\addplot [name path=lower_gg_L63_d2,draw=none] table[x=x,y expr=\thisrow{y}-\thisrow{err}] {gg_L63_t_d2.dat};
\addplot[
    color=green,
    mark=o,
    ]
    table[x=x, y=y]{gg_L63_t_d2.dat};
\addplot [fill=green!50, fill opacity=0.5] fill between[of=upper_gg_L63_d2 and lower_gg_L63_d2];
\addplot [name path=upper_pmle_L63_d2,draw=none] table[x=x,y expr=\thisrow{y}+\thisrow{err}] {pmle_L63_t_d2.dat};
\addplot [name path=lower_pmle_L63_d2,draw=none] table[x=x,y expr=\thisrow{y}-\thisrow{err}] {pmle_L63_t_d2.dat};
\addplot[
    color=blue,
    mark=diamond,
    ]
    table[x=x, y=y]{pmle_L63_t_d2.dat};
\addplot [fill=blue!50, fill opacity=0.5] fill between[of=upper_pmle_L63_d2 and lower_pmle_L63_d2];

\end{axis}
\end{tikzpicture}

%% file: plots_tex/plot_leg_L96.tex
\begin{tikzpicture}
\begin{axis}[
    legend style={font=\footnotesize, nodes={scale=0.85, transform shape}},
    hide axis,
    xmin=10,
    xmax=50,
    ymin=0,
    ymax=0.4,
    grid style=dashed,
    legend entries={B-GraphGrad, pMLE},
    legend columns = 2,
    height = 1.6in,
    width = 1.6in,
]
\addlegendimage{mark=o, color = green}
\addlegendimage{mark=diamond, color = blue}
\end{axis}
\end{tikzpicture}

%% file: plots_tex/L63_varT_d3.tex
\begin{tikzpicture}
\begin{axis}[
    xlabel={\footnotesize Series length},
    ylabel={\footnotesize RMSE},
    xmin=23.0, xmax=202.0,
    ymin=0.0, ymax=0.004,
    xtick={25, 50, 100, 200},
    ytick={0.0, 0.001, 0.002, 0.003, 0.004},
    legend style={at={(0.5,-0.3)},anchor=north,font=\footnotesize},
    ymajorgrids=true,
    grid style=dashed,
    ticklabel style = {font = \footnotesize},
    yticklabel style={/pgf/number format/fixed},
    height = 1.2in,
    width = 0.45\textwidth,
]


\addplot [name path=upper_gg_L63_d3,draw=none] table[x=x,y expr=\thisrow{y}+\thisrow{err}] {gg_L63_t_d3.dat};
\addplot [name path=lower_gg_L63_d3,draw=none] table[x=x,y expr=\thisrow{y}-\thisrow{err}] {gg_L63_t_d3.dat};
\addplot[
    color=green,
    mark=o,
    ]
    table[x=x, y=y]{gg_L63_t_d3.dat};
\addplot [fill=green!50, fill opacity=0.5] fill between[of=upper_gg_L63_d3 and lower_gg_L63_d3];
\addplot [name path=upper_pmle_L63_d3,draw=none] table[x=x,y expr=\thisrow{y}+\thisrow{err}] {pmle_L63_t_d3.dat};
\addplot [name path=lower_pmle_L63_d3,draw=none] table[x=x,y expr=\thisrow{y}-\thisrow{err}] {pmle_L63_t_d3.dat};
\addplot[
    color=blue,
    mark=diamond,
    ]
    table[x=x, y=y]{pmle_L63_t_d3.dat};
\addplot [fill=blue!50, fill opacity=0.5] fill between[of=upper_pmle_L63_d3 and lower_pmle_L63_d3];

\end{axis}
\end{tikzpicture}

%% file: plots_tex/L36_varsigma2_d2.tex
\begin{tikzpicture}
\pgfplotsset{%
    x tick label style={/pgf/number format/1000 sep=\,},
    log base 10 number format code/.code={%
        $\pgfmathparse{10^(#1)}\pgfmathprintnumber{\pgfmathresult}$%
    }%
  }  
\begin{axis}[
    xlabel={\footnotesize $\sigma^2$},
    ylabel={\footnotesize RMSE},
    xmin=0.009, xmax=5.5,
    ymin=0.0, ymax=0.0025,
    xtick={0.01, 0.1, 1, 5},
    ytick={0.0, 0.0005, 0.001, 0.0015, 0.002, 0.0025},
    legend style={at={(0.5,-0.3)},anchor=north,font=\footnotesize},
    ymajorgrids=true,
    grid style=dashed,
    xmode=log,
    ticklabel style = {font = \footnotesize},
    yticklabel style={/pgf/number format/fixed},
    height = 1.2in,
    width = 0.45\textwidth,
]


\addplot [name path=upper_gg_L63_d2_s2,draw=none] table[x=x,y expr=\thisrow{y}+\thisrow{err}] {gg_L63_s2_d2.dat};
\addplot [name path=lower_gg_L63_d2_s2,draw=none] table[x=x,y expr=\thisrow{y}-\thisrow{err}] {gg_L63_s2_d2.dat};
\addplot[
    color=green,
    mark=o,
    ]
    table[x=x, y=y]{gg_L63_s2_d2.dat};
\addplot [fill=green!50, fill opacity=0.5] fill between[of=upper_gg_L63_d2_s2 and lower_gg_L63_d2_s2];
\addplot [name path=upper_pmle_L63_d2_s2,draw=none] table[x=x,y expr=\thisrow{y}+\thisrow{err}] {pmle_L63_s2_d2.dat};
\addplot [name path=lower_pmle_L63_d2_s2,draw=none] table[x=x,y expr=\thisrow{y}-\thisrow{err}] {pmle_L63_s2_d2.dat};
\addplot[
    color=blue,
    mark=diamond,
    ]
    table[x=x, y=y]{pmle_L63_s2_d2.dat};
\addplot [fill=blue!50, fill opacity=0.5] fill between[of=upper_pmle_L63_d2_s2 and lower_pmle_L63_d2_s2];

\end{axis}
\end{tikzpicture}

%% file: plots_tex/L96_varT_d2.tex
\begin{tikzpicture}
\begin{axis}[
    xlabel={\footnotesize Series length},
    ylabel={\footnotesize RMSE},
    xmin=23.0, xmax=202.0,
    ymin=0.0, ymax=0.004,
    xtick={25, 50, 100, 200},
    ytick={0.0, 0.001, 0.002, 0.003, 0.004},
    legend style={at={(0.5,-0.3)},anchor=north,font=\footnotesize},
    ymajorgrids=true,
    grid style=dashed,
    ticklabel style = {font = \footnotesize},
    yticklabel style={/pgf/number format/fixed},
    height = 1.2in,
    width = 0.45\textwidth,
]


\addplot [name path=upper_gg_L96_d2,draw=none] table[x=x,y expr=\thisrow{y}+\thisrow{err}] {gg_L96_t_d2.dat};
\addplot [name path=lower_gg_L96_d2,draw=none] table[x=x,y expr=\thisrow{y}-\thisrow{err}] {gg_L96_t_d2.dat};
\addplot[
    color=green,
    mark=o,
    ]
    table[x=x, y=y]{gg_L96_t_d2.dat};
\addplot [fill=green!50, fill opacity=0.5] fill between[of=upper_gg_L96_d2 and lower_gg_L96_d2];
\addplot [name path=upper_pmle_L96_d2,draw=none] table[x=x,y expr=\thisrow{y}+\thisrow{err}] {pmle_L96_t_d2.dat};
\addplot [name path=lower_pmle_L96_d2,draw=none] table[x=x,y expr=\thisrow{y}-\thisrow{err}] {pmle_L96_t_d2.dat};
\addplot[
    color=blue,
    mark=diamond,
    ]
    table[x=x, y=y]{pmle_L96_t_d2.dat};
\addplot [fill=blue!50, fill opacity=0.5] fill between[of=upper_pmle_L96_d2 and lower_pmle_L96_d2];

\end{axis}
\end{tikzpicture}

%% file: plots_tex/L96_varT_d3.tex
\begin{tikzpicture}
\begin{axis}[
    xlabel={\footnotesize Series length},
    ylabel={\footnotesize RMSE},
    xmin=23.0, xmax=202.0,
    ymin=0.0, ymax=0.004,
    xtick={25, 50, 100, 200},
    ytick={0.0, 0.001, 0.002, 0.003, 0.004},
    legend style={at={(0.5,-0.3)},anchor=north,font=\footnotesize},
    ymajorgrids=true,
    grid style=dashed,
    ticklabel style = {font = \footnotesize},
    yticklabel style={/pgf/number format/fixed},
    height = 1.2in,
    width = 0.45\textwidth,
]


\addplot [name path=upper_gg_L96_d3,draw=none] table[x=x,y expr=\thisrow{y}+\thisrow{err}] {gg_L96_t_d3.dat};
\addplot [name path=lower_gg_L96_d3,draw=none] table[x=x,y expr=\thisrow{y}-\thisrow{err}] {gg_L96_t_d3.dat};
\addplot[
    color=green,
    mark=o,
    ]
    table[x=x, y=y]{gg_L96_t_d3.dat};
\addplot [fill=green!50, fill opacity=0.5] fill between[of=upper_gg_L96_d3 and lower_gg_L96_d3];
\addplot [name path=upper_pmle_L96_d3,draw=none] table[x=x,y expr=\thisrow{y}+\thisrow{err}] {pmle_L96_t_d3.dat};
\addplot [name path=lower_pmle_L96_d3,draw=none] table[x=x,y expr=\thisrow{y}-\thisrow{err}] {pmle_L96_t_d3.dat};
\addplot[
    color=blue,
    mark=diamond,
    ]
    table[x=x, y=y]{pmle_L96_t_d3.dat};
\addplot [fill=blue!50, fill opacity=0.5] fill between[of=upper_pmle_L96_d3 and lower_pmle_L96_d3];

\end{axis}
\end{tikzpicture}

%% file: plots_tex/kura_varT.tex
\begin{tikzpicture}
\begin{axis}[
    xlabel={\footnotesize Series length},
    ylabel={\footnotesize nRMSE},
    xmin=23.0, xmax=202.0,
    ymin=0.95, ymax=1.35,
    xtick={25, 50, 100, 200},
    ytick={1, 1.1, 1.2, 1.3},
    legend style={at={(0.5,-0.3)},anchor=north,font=\footnotesize},
    ymajorgrids=true,
    grid style=dashed,
    ticklabel style = {font = \footnotesize},
    height = 1.2in,
    width = 0.45\textwidth,
]

\addplot [color=black,dashed] coordinates {(0,1) (220,1)};

\addplot [name path=upper_gg_kura,draw=none] table[x=x,y expr=\thisrow{y}+\thisrow{err}] {gg_kura_t.dat};
\addplot [name path=lower_gg_kura,draw=none] table[x=x,y expr=\thisrow{y}-\thisrow{err}] {gg_kura_t.dat};
\addplot[
    color=green,
    mark=o,
    ]
    table[x=x, y=y]{gg_kura_t.dat};
\addplot [fill=green!50, fill opacity=0.5] fill between[of=upper_gg_kura and lower_gg_kura];
\addplot [name path=upper_pmle_kura,draw=none] table[x=x,y expr=\thisrow{y}+\thisrow{err}] {pmle_kura_t.dat};
\addplot [name path=lower_pmle_kura,draw=none] table[x=x,y expr=\thisrow{y}-\thisrow{err}] {pmle_kura_t.dat};
\addplot[
    color=blue,
    mark=diamond,
    ]
    table[x=x, y=y]{pmle_kura_t.dat};
\addplot [fill=blue!50, fill opacity=0.5] fill between[of=upper_pmle_kura and lower_pmle_kura];
\addplot [name path=upper_truemle_kura,draw=none] table[x=x,y expr=\thisrow{y}+\thisrow{err}] {truemle_kura_t.dat};
\addplot [name path=lower_truemle_kura,draw=none] table[x=x,y expr=\thisrow{y}-\thisrow{err}] {truemle_kura_t.dat};
\addplot[
    color=red,
    mark=square,
    ]
    table[x=x, y=y]{truemle_kura_t.dat};
\addplot [fill=red!50, fill opacity=0.5] fill between[of=upper_truemle_kura and lower_truemle_kura];

\end{axis}
\end{tikzpicture}

%% file: plots_tex/plot_leg_kura.tex
\begin{tikzpicture}
\begin{axis}[
    legend style={font=\footnotesize, nodes={scale=0.85, transform shape}},
    hide axis,
    xmin=10,
    xmax=50,
    ymin=0,
    ymax=0.4,
    grid style=dashed,
    legend entries={B-GraphGrad, pMLE, TrueMLE},
    legend columns = 3,
    height = 1.6in,
    width = 1.6in,
]
\addlegendimage{mark=o, color = green}
\addlegendimage{mark=diamond, color = blue}
\addlegendimage{mark=square, color = red}
\addlegendimage{color = black, dashed}
\end{axis}
\end{tikzpicture}